\documentclass[twocolumn, 10pt, aps, superscriptaddress, floatfix, showpacs, prb, citeautoscript]{revtex4-1}

\usepackage[utf8]{inputenc}
\usepackage[T1]{fontenc}
\usepackage{graphicx}
\usepackage{amsmath}
\usepackage{amssymb}
\usepackage{bm}
\usepackage{xcolor}
\usepackage[colorlinks, citecolor={blue!50!black}, urlcolor={blue!50!black}, linkcolor={red!50!black}]{hyperref}
\usepackage{bookmark}
\usepackage{tabularx}
\usepackage{mathtools}
\usepackage{microtype}
\usepackage[load=physical,load=abbr]{siunitx}

\newcommand{\co}[2]{#2}

\DeclarePairedDelimiter\abs{\lvert}{\rvert}%
\DeclarePairedDelimiter\norm{\lVert}{\rVert}%
\makeatletter
\let\oldabs\abs
\def\abs{\@ifstar{\oldabs}{\oldabs*}}
\let\oldnorm\norm
\def\norm{\@ifstar{\oldnorm}{\oldnorm*}}
\makeatother

\newcolumntype{L}[1]{>{\raggedright\arraybackslash}p{#1}}
\newcolumntype{C}[1]{>{\centering\arraybackslash}p{#1}}
\newcolumntype{R}[1]{>{\raggedleft\arraybackslash}p{#1}}

\graphicspath{{figures/}}

\begin{document}

\title{Effects of the electrostatic environment on the Majorana nanowire devices}

\author{A. Vuik}
\email[Electronic address: ]{adriaanvuik@gmail.com}
\affiliation{Kavli Institute of Nanoscience, Delft University of Technology,
  P.O. Box 4056, 2600 GA Delft, The Netherlands}
\author{D. Eeltink}
\altaffiliation[Current address: ]{Universit\'e de Gen\`eve, GAP-Biophotonics, Chemin de Pinchat 22, CH-1211 Geneva 4, Switzerland}
\affiliation{Kavli Institute of Nanoscience, Delft University of Technology,
  P.O. Box 4056, 2600 GA Delft, The Netherlands}
\author{A. R. Akhmerov}
\affiliation{Kavli Institute of Nanoscience, Delft University of Technology,
  P.O. Box 4056, 2600 GA Delft, The Netherlands}
\author{M. Wimmer}
\affiliation{QuTech, Delft University of Technology,
  P.O. Box 4056, 2600 GA Delft, The Netherlands}
\affiliation{Kavli Institute of Nanoscience, Delft University of Technology,
  P.O. Box 4056, 2600 GA Delft, The Netherlands}

\date{25 November 2015}

\begin{abstract}
One of the promising platforms for creating Majorana bound states is a hybrid nanostructure consisting of a semiconducting nanowire covered by a superconductor.
We analyze the previously disregarded role of electrostatic interaction in these devices.
Our main result is that Coulomb interaction causes the chemical potential to respond to an applied magnetic field, while spin-orbit interaction and screening by the superconducting lead suppress this response.
Consequently, the electrostatic environment influences two properties of Majorana devices: the shape of the topological phase boundary and the oscillations of the Majorana splitting energy.
We demonstrate that both properties show a non-universal behavior, and depend on the details of the electrostatic environment.
We show that when the wire only contains a single electron mode, the experimentally accessible inverse self-capacitance of this mode fully captures the interplay between electrostatics and Zeeman field.
This offers a way to compare theoretical predictions with experiments.
\end{abstract}

\maketitle
\section{Introduction}

\co{Realization of Majoranas as bound states in hybrid superconductor-semiconductor devices}

Majorana zero modes are non-Abelian anyons that emerge in condensed-matter systems as zero-energy excitations in superconductors~\cite{Fu2008, Alicea2012, Beenakker2013}.
They exhibit non-Abelian braiding statistics \cite{Alicea2011} and form a building block for topological quantum computation~\cite{Nayak2008}.
Following theoretical proposals\cite{Oreg2010, Lutchyn2010}, experiments in semiconducting nanowires with proximitized superconductivity report appearance of Majorana zero modes signatures~\cite{Mourik2012, Das2012, Deng2012, Churchill2013, Deng2014}.
These ``Majorana devices'' are expected to switch from a trivial to a topological state when a magnetic field closes the induced superconducting gap.
A further increase of the magnetic field reopens the bulk gap again with Majorana zero modes remaining at the edges of the topological phase.

\co{Screening by the superconductor can be too strong to control the electron density}

Inducing superconductivity requires close proximity of the nanowire to a superconductor, which screens the electric field created by gate voltages.
Another source of screening is the charge in the nanowire itself that counteracts the applied electric field.
Therefore, a natural concern in device design is whether these screening effects prevent effective gating of the device.
Besides this, screening effects and work function differences between the superconductor and the nanowire affect the spatial structure of the electron density in the wire.
The magnitude of the induced superconducting gap reduces when charge localizes far away from the superconductor, restricting the parameter range for the observation of Majorana modes.

\co{We study influence of the electrostatic environment}

To quantitatively assess these phenomena, we study the influence of the electrostatic environment on the properties of Majorana devices.
We investigate the effect of screening by the superconductor as a function of the work function difference between the superconductor and the nanowire, and we study screening effects due to charge.
We focus on the influence of screening on the behavior of the chemical potential in the presence of a magnetic field in particular, because the chemical potential directly impacts the Majorana signatures.

\co{This is also relevant for Majorana signatures}

The zero-bias peak, measured experimentally in Refs.~\onlinecite{Mourik2012, Das2012, Deng2012, Churchill2013, Deng2014}, is a non-specific signature of Majoranas, since similar features arise due to Kondo physics or weak anti-localization \cite{Bagrets2012, Pikulin2012}.
To help distinguishing Majorana signatures from these alternatives, we focus on the parametric dependence of two Majorana properties: the shape of the topological phase boundary \cite{Stanescu2012, Fregoso2013} and the oscillations in the coupling energy of two Majorana modes \cite{Cheng2009, Cheng2010, Prada2012, Sarma2012, Rainis2013}.

\co{The two proposed Majorana signatures depend on mu and thus on electrostatics}

Both phenomena depend on the response of the chemical potential to a magnetic field, and hence on electrostatic effects. Majorana oscillations were analyzed theoretically in two extreme limits for the electrostatic effects: constant chemical potential \cite{Prada2012, Sarma2012, Rainis2013} and constant density \cite{Sarma2012} (see App.~\ref{app:nomenclature} for a summary of these two limits). In particular, Ref.~\onlinecite{Sarma2012} found different behavior of Majorana oscillations in these two extreme limits. We show that the actual behavior of the nanowire is somewhere in between, and depends strongly on the electrostatics.

\section{Setup and methods}
\label{sec:SP}

\subsection{The Schr\"odinger-Poisson problem}

\co{Description of the system and boundary conditions}

\begin{figure}[tb]
\centerline{\includegraphics[width=1.0\linewidth]{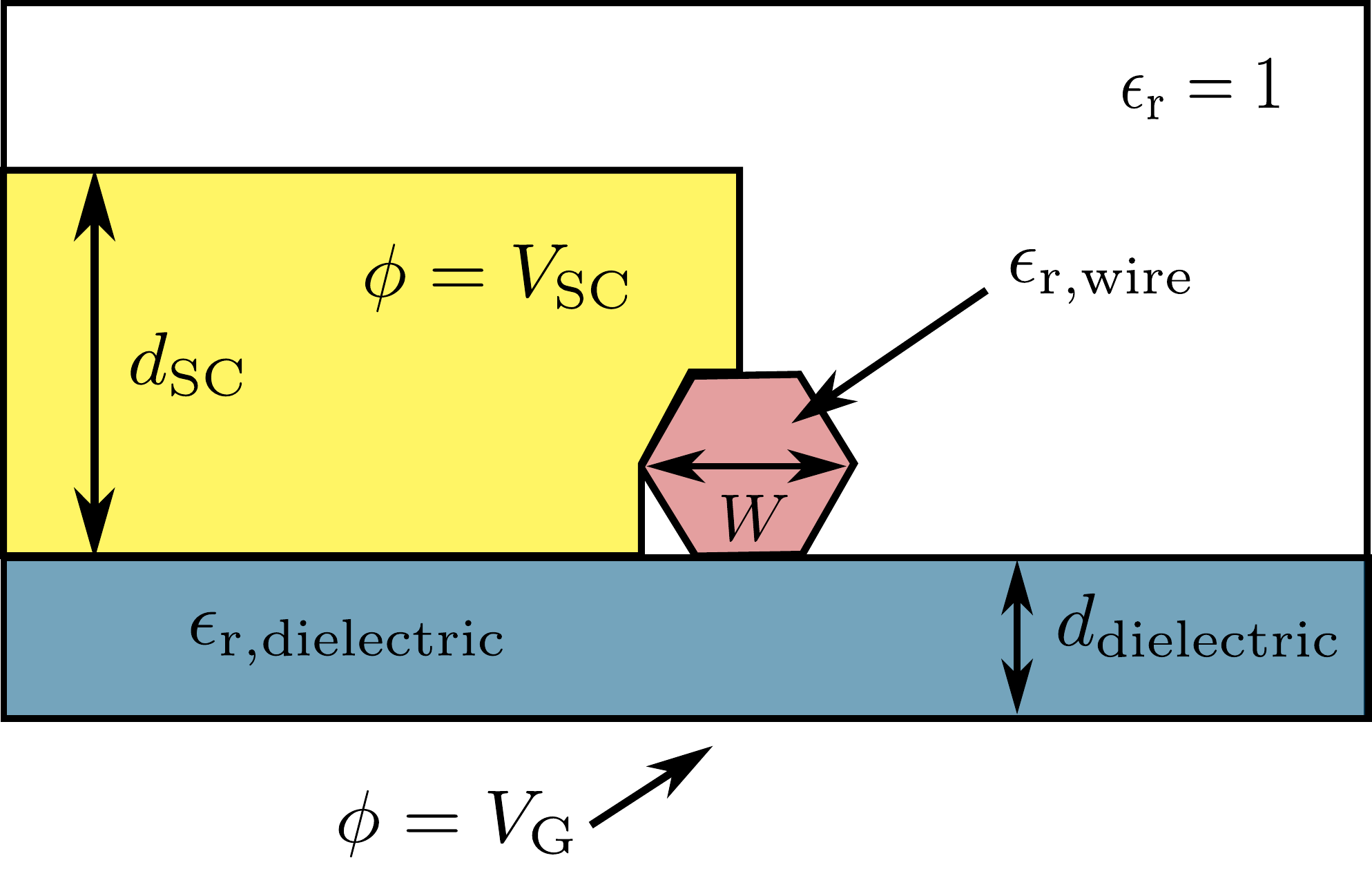}}
\caption{Schematic cross section of the Majorana device.
It consists of a nanowire (red hexagon) lying on a  dielectric layer (blue rectangle) which covers a global back gate. A superconducting lead (yellow region) covers half of the nanowire.}
\label{fig:boundary_cond}
\end{figure}

We discuss electrostatic effects in a device design as used by Mourik et al~\cite{Mourik2012}, however our methods are straightforward to adapt to similar layouts (see App.~\ref{app:marcus_device} for a calculation using a different geometry).
Since we are interested in the bulk properties, we require that the potential and the Hamiltonian terms are translationally invariant along the wire axis and we consider a 2D cross section, shown in Fig.~\ref{fig:boundary_cond}.
The device consists of a nanowire with a hexagonal cross section of diameter $W = \SI{100}{\nm}$ on a dielectric layer with thickness $d_{\textrm{dielectric}} =\SI{30}{\nano\meter}$. A superconductor with thickness $d_{\textrm{SC}}= \SI{187}{\nano\meter}$ covers half of the wire. The nanowire has a dielectric constant $\epsilon_{\textrm{r}} = 17.7$ (InSb), the dielectric layer has a dielectric constant $\epsilon_{\textrm{r}} = 8$ (Si$_3$N$_4$).
The device has two electrostatic boundary conditions: a fixed gate potential $V_{\textrm{G}}$ set by the gate electrode along the the lower edge of the dielectric layer and a fixed potential $V_{\textrm{SC}}$ in the superconductor, which we model as a grounded metallic gate.
We set this potential to either $V_{\textrm{SC}} = \SI{0}{\volt}$, disregarding a work function difference between the NbTiN superconductor and the nanowire, or we assume a small work function difference \cite{Haneman1959, Wilson1966} resulting in $V_{\textrm{SC}} = \SI{0.2}{\volt}$.

\co{The Schr\"odinger equation}

\begin{figure}
\includegraphics[width=\linewidth]{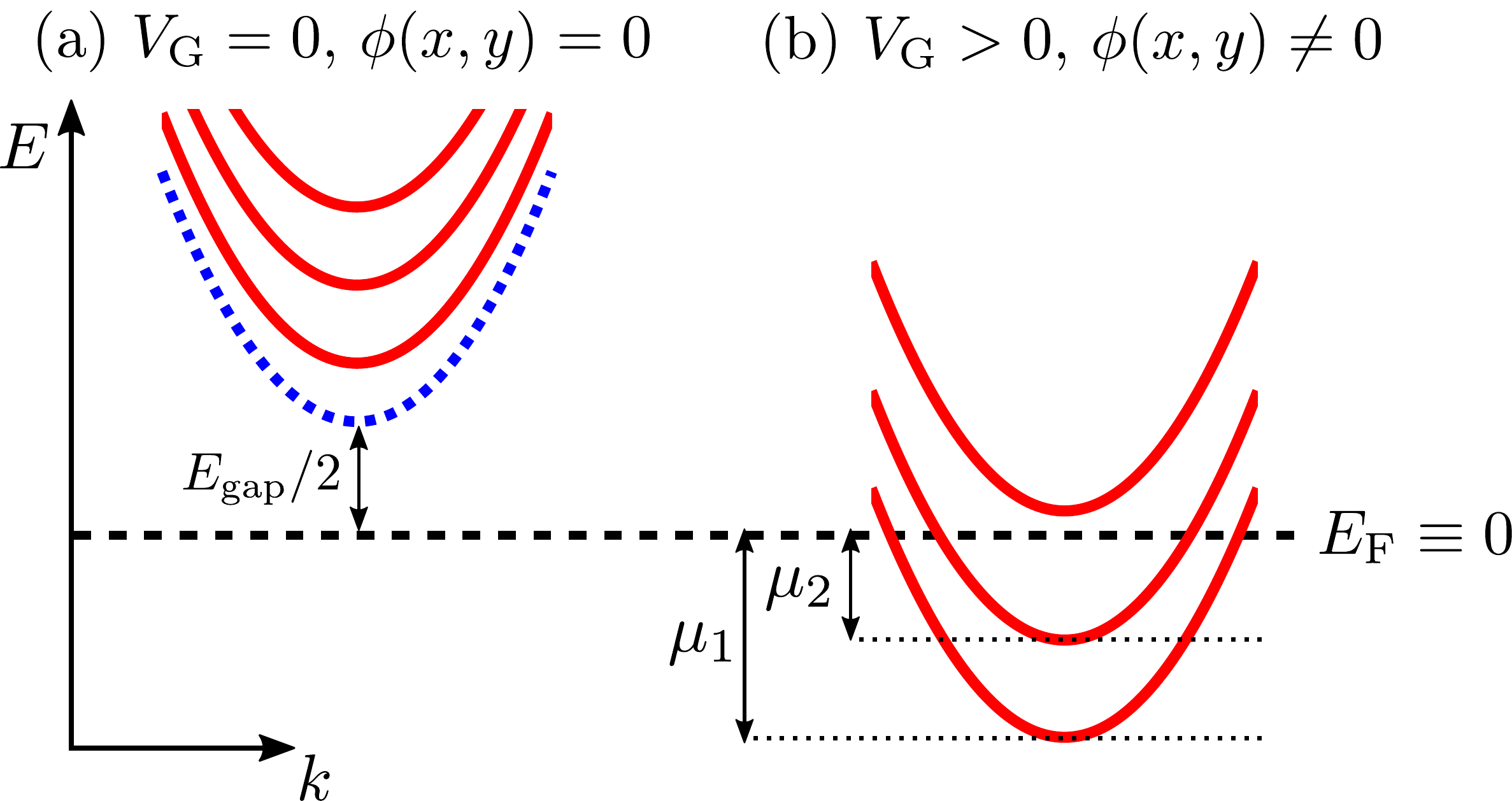}
\caption{Band alignment and the Fermi level, shown schematically for
  $V_\text{SC}=0$.
  (a) In the absence of an electrostatic potential (gate voltage $V_\text{G}=0$) the Fermi level $E_\text{F}$ is assumed to be aligned to the middle of the semiconducting gap (of
  size $E_\text{gap}$, semiconductor conduction band shown as dashed blue line).
  Confinement in the nanowire leads to discrete subbands (red solid lines).
  (b) A positive gate voltage gives rise to an electrostatic potential landscape lowering the energy of all subbands.
  Subbands below the Fermi level $E_\text{F}$ are occupied.
  For those bands, we define effective chemical potentials $\mu_i$.
  (Note that the subband spacings depend on $\phi(x,y)$ and are typically different for different $V_\text{G}$.)
  For simplicity, we set the spin-orbit interaction to zero in these dispersions.
  For nonzero spin-orbit strength, the chemical potentials $\mu_i$ are defined with respect to the crossing point of the spin bands rather than at the band edge.}\label{fig:band_align}
\end{figure}

We model the electrostatics of this setup using the Schr\"odinger-Poisson equation.
We split the Hamiltonian into transverse and longitudinal parts.
The transverse Hamiltonian $\mathcal{H}_{\textrm{T}}$ reads
\begin{equation}
\mathcal{H}_{\textrm{T}} = -\frac{\hbar^2}{2m^*}\left( \frac{\partial^2}{\partial x^2} + \frac{\partial^2}{\partial y^2} \right) - e \phi(x, y) + \frac{E_\text{gap}}{2},
\label{eq:transverseHamiltonian}
\end{equation}
with $x, y$ the transverse directions, $m^* = 0.014 m_e$ the effective
electron mass in InSb (with $m_e$ the electron mass), $-e$ the electron charge, and $\phi$ the electrostatic potential.
We assume that in the absence of electric field the Fermi level
$E_\text{F}$ in the nanowire is in the middle of the semiconducting gap
$E_\text{gap}$, with $E_\text{gap} = \SI{0.2}{\eV}$ for InSb (see Fig.~\ref{fig:band_align}(a).
We choose the Fermi level $E_\text{F}$ as the reference energy such that $E_\text{F} \equiv 0$.

The longitudinal Hamiltonian $\mathcal{H}_{\textrm{L}}$ reads
\begin{equation}
\mathcal{H}_{\textrm{L}} = -\frac{\hbar^2}{2m^*}\frac{\partial^2}{\partial z^2} - i \alpha \frac{\partial}{\partial z} \sigma_y + E_{\textrm{Z}} \sigma_z,
\label{eq:longHamiltonian}
\end{equation}
with $z$ the direction along the wire axis, $\alpha$ the spin-orbit coupling strength, $E_{\textrm{Z}}$ the Zeeman energy and $\bm{\sigma}$ the Pauli matrices. The orientation of the magnetic field is along the wire in the $z$ direction.
In this separation, we have assumed that the spin-orbit length $l_\textrm{SO}=\hbar^2/(m^* \alpha)$ is larger or comparable to the wire diameter,
$l_\text{SO} \gtrsim W$. \cite{Scheid2009, Diez2012} Furthermore, we neglect the explicit dependence of the spin-orbit strength $\alpha$ on the electric field. We ignore orbital effects of the magnetic field, \cite{Nijholt2015} since the effective area of the transverse wave functions is much smaller than the wire cross section due to screening by the superconductor, as we show in Sec. \ref{sec:screeningeffects}.

\co{Coupling between Schrodinger and Poisson: the density}

Since the Hamiltonian is separable in the limit we are using, the charge density in the transverse direction $\rho(x, y)$ is:
\begin{equation}
\rho(x,y) = -e\sum_i|\psi_i(x, y)|^2 \ n(E_i, E_{\textrm{Z}}, \alpha),
\label{eq:density}
\end{equation}
with $\psi_i$ the transverse wave function and $E_i$ the subband energy of the $i$-th electron mode defined by $\mathcal{H}_\text{T} \psi_i = E_i \psi_i$.
Further, $n(E_i, E_\text{Z}, \alpha)$ is the 1D electron density, which we calculate in closed form from the Fermi momenta of different bands in App.~\ref{app:density}. The subband energies $E_i$ depend on the electrostatic potential $\phi(x,y)$,
and individual subbands are occupied by ``lowering'' subbands below $E_\text{F}$
(shown schematically in Fig.~\ref{fig:band_align}(b)). \footnote{Note that $E_i$ agrees with the subband bottom only if $\alpha=0$ and $E_\text{Z}$=0. See App.~\ref{app:density} for details on the subband occupation in the general case.}

\co{The Poisson equation}

The Poisson equation that determines the electrostatic potential $\phi(x, y)$ has the general form:
\begin{equation}
\nabla^2 \phi(x, y) = -\frac{\rho(x,y)}{\epsilon},
\label{eq:Poisson}
\end{equation}
with $\epsilon$ the dielectric permittivity.
Since the charge density of Eq.~\eqref{eq:density} depends on the eigenstates of
Eq.~\eqref{eq:transverseHamiltonian}, the Schr\"odinger and the Poisson
equations have a nonlinear coupling.

\co{Numerical procedure for solving SP}

We calculate the eigenstates and eigenenergies of the Hamiltonian of Eq.~\eqref{eq:transverseHamiltonian} in tight-binding approximation on a rectangular grid using the Kwant package \cite{Groth2014}.
We then discretize the geometry of Fig.~\ref{fig:boundary_cond} using a finite element mesh, and solve Eq.~\eqref{eq:Poisson} numerically using the FEniCS package \cite{Logg2012}.

\co{We solve self-consistently using the Anderson method}

Eqs.~\eqref{eq:transverseHamiltonian} and \eqref{eq:density} together define a
functional $\bar{\rho}[\phi]$, yielding a charge density from a given
electrostatic potential $\phi$. Additionally, Eq.~\eqref{eq:Poisson}
defines a functional $\bar{\phi}[\rho]$, giving the electrostatic potential
produced by a charge density $\rho$. The Schr\"odinger-Poisson equation is self-consistent when
\begin{equation}
\bar{\phi}[\bar{\rho}[\phi]] - \phi = 0.
\label{eq:minimize_func}
\end{equation}
We solve Eq.~\eqref{eq:minimize_func} using an iterative nonlinear Anderson mixing method \cite{Eyert1996}.
We find that this method prevents the iteration process from oscillations and leads to a significant speedup in computation times compared to other nonlinear solver methods (see App.~\ref{app:benchmarks}).
We search for the root of Eq.~\eqref{eq:minimize_func} rather than for the root of
\begin{equation}
\bar{\rho}[\bar{\phi}[\rho]] - \rho = 0,
\label{eq:minimize_func_rho}
\end{equation}
since we found Eq.~\eqref{eq:minimize_func} to be better conditioned than Eq.~\eqref{eq:minimize_func_rho}. The scripts with the source code as well as resulting data are available online as ancillary files for this manuscript.

\subsection{Majorana zero modes in superconducting nanowires}\label{sec:majoranamodestheory}

Having solved the electrostatic problem for the normal system, i.e.~taking into account only the electrostatic effects of the superconductor, we then use the electrostatic potential $\phi(x,y)$ in the superconducting problem.
To this end, we obtain the Bogoliubov-de Gennes Hamiltonian $\mathcal{H}_{\textrm{BdG}}$ by summing $\mathcal{H}_\text{T}$ and $\mathcal{H}_\text{L}$ and adding an induced superconducting pairing term:
\begin{multline}
\mathcal{H}_{\textrm{BdG}} = \\ \left[ \left( -\frac{\hbar^2}{2m^*}\nabla^2 -e \phi(x,y) + \frac{E_{gap}}{2} \right) \sigma_0  - i \alpha \frac{\partial}{\partial z} \sigma_y \right] \otimes \tau_z  \\
+ E_{\textrm{Z}} \sigma_z \otimes \tau_0 + \Delta \sigma_0 \otimes \tau_x,
\label{eq:BdGHamiltonian}
\end{multline}
with $\bm{\tau}$ the Pauli matrices in electron-hole space and $\Delta$ the superconducting gap.

The three-dimensional BdG equation \eqref{eq:BdGHamiltonian} is still
separable and reduces for every subband with transverse wave function
$\psi_i$ to an effective one-dimensional BdG Hamiltonian:
\begin{multline}
\mathcal{H}_{\textrm{BdG},i} = \left[ \left( \frac{p^2}{2m^*} - \mu_i \right) \sigma_0  + \frac{\alpha}{\hbar} p \sigma_y \right] \otimes \tau_z  \\
+ E_{\textrm{Z}} \sigma_z \otimes \tau_0 + \Delta \sigma_0 \otimes \tau_x,
\label{eq:BdGHamiltonian1D}
\end{multline}
where $p = -i\hbar\, \partial/\partial z$ and we defined $\mu_i = -E_i$
(see Fig.~\ref{fig:band_align}(b). Since the different subbands are independent,
$\mu_i$ can be interpreted as the chemical potential determining the
occupation of the $i$-th subband.

While the Fermi level is kept constant by the metallic contacts, the chemical potential $\mu_i$ of each subband does depend on the system parameters: $\mu_i=\mu_i(V_\text{G}, E_\text{Z})$.
Most of the model Hamiltonians for Majorana nanowires used in the literature are of the form of Eq.~\eqref{eq:BdGHamiltonian1D} (or a two-dimensional generalization) using one chemical potential $\mu$. 
To make the connection to our work, $\mu$ should be identified with $\mu_i$, and not be confused with the constant Fermi level $E_\text{F}$.
For example, the constant chemical potential limit of Ref.~\onlinecite{Sarma2012} refers to the special case that $\mu_i$ is independent of $E_\text{Z}$, and it is not related to $E_\text{F}$ being always constant. \footnote{Using the notion of a variable chemical potential $\mu$ is natural when energies are measured with respect to a fixed band bottom, i.e. in a single-band situation. For us,
different subbands react differently on changes in $\phi(x,y)$ and it is more practical to keep the Fermi level $E_\text{F}$ fixed.}

Properties of Majorana modes formed in the $i$-th subband only depend on the value of $\mu_i$ (or equivalently $E_i$).
In the following we thus determine the effect of the electrostatics on $\mu_i$ before we finally turn to Majorana bound states.

\section{Screening effects on charge density and energy levels}
\label{sec:screeningeffects}

\co{We calculate dispersions and charge densities turning different types of screening on and off.}

\begin{figure}[tbh]
\centerline{\includegraphics[width=.9\linewidth]{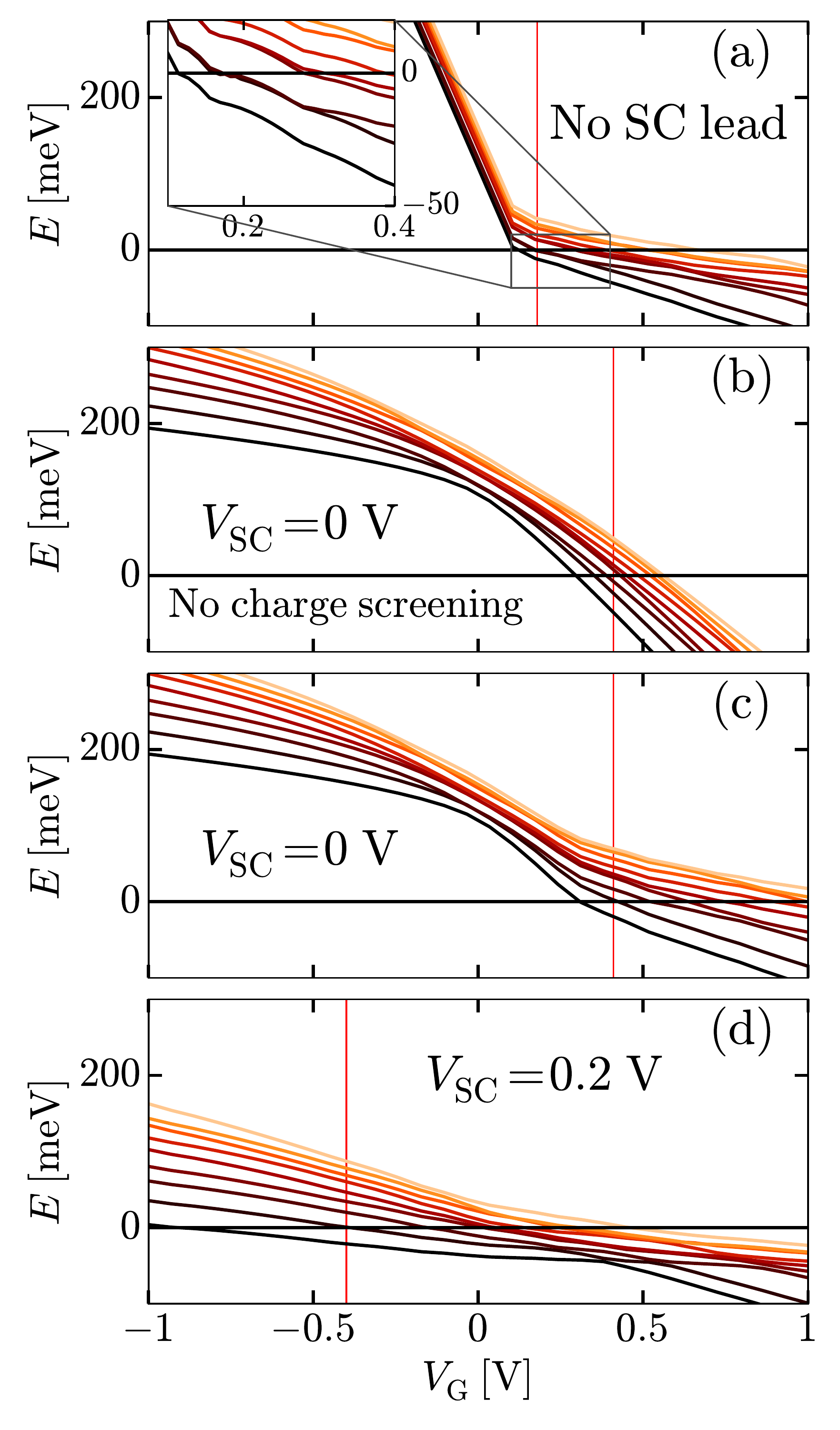}}
\caption{The nine lowest subband energies $\mu_i$ as a function of gate voltage. (a): Wire without a superconducting lead, (b): wire with a superconducting lead at $V_{\textrm{SC}} = \SI{0}{\volt}$, neglecting charge screening effects, (c): the same problem including charge screening effects, and (d): a superconducting lead with $V_{\textrm{SC}} = \SI{0.2}{\volt}$ including charge screening.
The Fermi level $E_F=0$ is indicated as a solid horizontal line.
The red lines indicates the gate voltages used in the calculation of charge density and electric field of the corresponding panels in Fig.~\ref{fig:screening}. In all plots, we take weak spin-orbit interaction (a spin-orbit length of \SI{233}{\nm}).
The inset of the top panel shows a zoom, revealing Fermi level pinning every time a new band crosses the Fermi level.}
\label{fig:energy_levels}
\end{figure}

\begin{figure}[tbh]
\centerline{\includegraphics[width=1.0\linewidth]{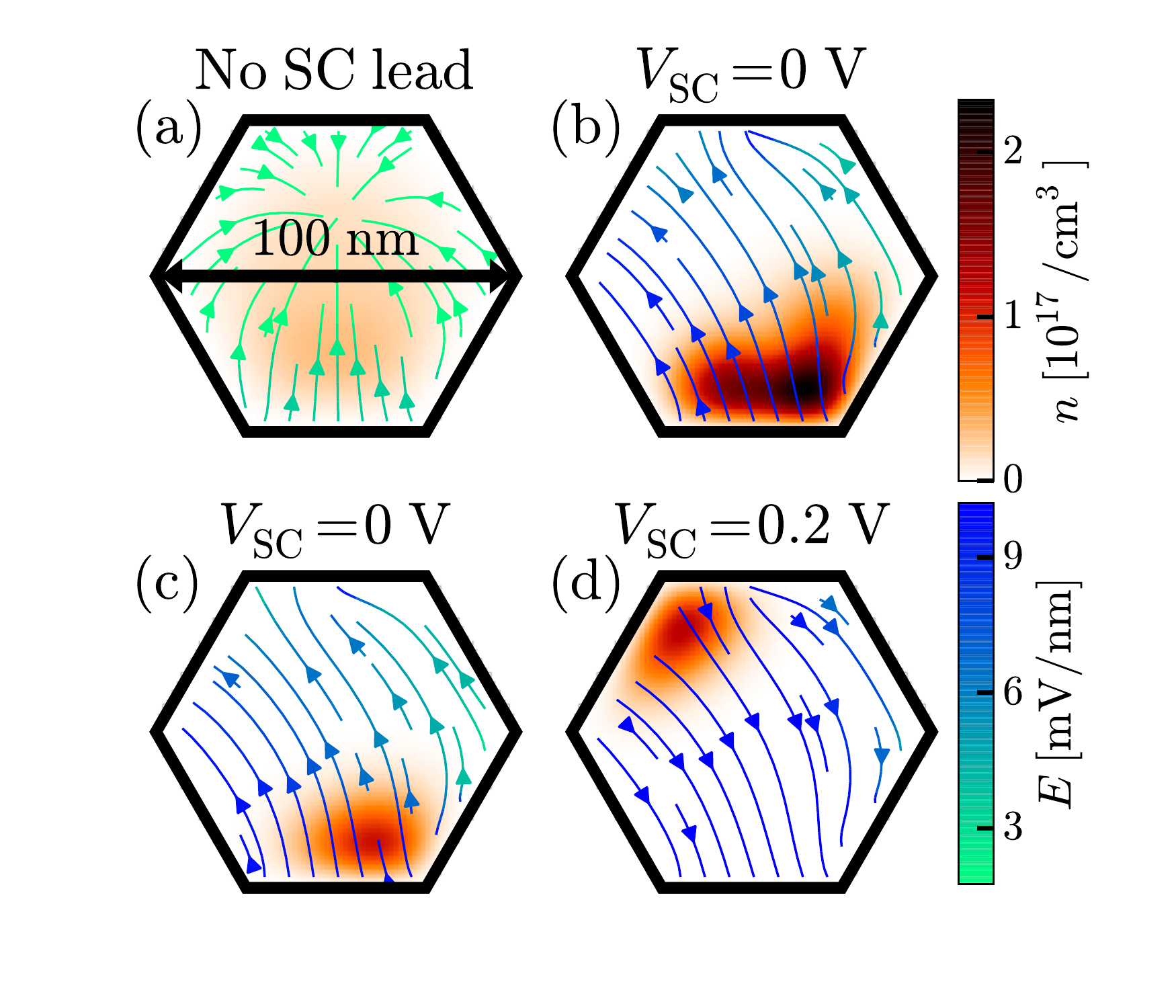}}
\caption{Charge density distribution and electric field in the wire cross section, at the gate voltage indicated by the red line in the corresponding panel of Fig.~\ref{fig:energy_levels}.
(a): Self-consistent solution when no superconducting lead is attached.
(b): Superconducting lead at $V_{\textrm{SC}} = \SI{0}{\volt}$, neglecting screening by charge.
(c): Same problem, but including screening by charge (self-consistent).
(d): Self-consistent solution for a superconducting lead at $V_{\textrm{SC}} = \SI{0.2}{\volt}$.
The total density is $\approx \SI{5.5e5}{\per \centi\meter}$ for plots (a), (c), and (d).
Plot (b) has a total density of $\approx \SI{1.6e6}{ \per \centi\meter}$.}
\label{fig:screening}
\end{figure}

We begin by investigating the electrostatic effects in absence of Zeeman field and a spin-orbit strength with $l_\textrm{SO}=\SI{233}{\nm}$, negligible for the electrostatic effects.
We solve the Schr\"odinger-Poisson equation for a superconductor with $V_{\textrm{SC}} = \SI{0}{\volt}$ and a superconductor with $V_{\textrm{SC}} = \SI{0.2}{\volt}$, and compare the solutions to two benchmarks: a nanowire without a superconducting lead, and a nanowire in which we ignore screening by charge.
Specifically, we compute the influence of screening by the superconductor and by charge on the field effect on the lowest energy levels and charge densities.
To evaluate the role of screening by charges in the wire, we compare the full solutions of the Poisson equation \eqref{eq:Poisson} to its solution with the right-hand side set to zero.
Our results are summarized in Fig.~\ref{fig:energy_levels} showing the dispersion of $\mu_i$ and Fig.~\ref{fig:screening} showing the charge density for the same situations and the values of $V_\textrm{G}$ marked in Fig.~\ref{fig:energy_levels}.

\co{The levels are almost degenerate without electric field.}
The approximate rotational symmetry of the wire leads to almost doubly degenerate bands with opposite angular momenta when electric field is negligible---a situation realized either in absence of the superconductor [Fig.~\ref{fig:energy_levels}(a)] or when $V_\textrm{G}=V_\textrm{SC}$ [Fig.~\ref{fig:energy_levels}(b), (c), (d)].
However in most cases, presence of the superconductor leads to a large $V_\textrm{G}$ required to induce a finite charge density in the wire, and the degeneracy is strongly lifted.

\co{Various types of screening have different impact on the energies of the levels.}
The lever arm of the gate voltage on the energies $E_i$, reduces from the optimal value of \SI{1}{}, at $V_\textrm{G} < 0$ by approximately a factor of \SI{4}{} due to charge screening alone [Fig.~\ref{fig:energy_levels}(a)].
Screening by the superconductor leads to an additional comparable suppression of the lever arm, however its effect is nonlinear in $V_\textrm{G}$ due to the transverse wave functions being pulled closer to the gate at positive $V_\textrm{G}$.
Comparing panels (b) and (c) of Fig.~\ref{fig:energy_levels} we see that screening by the superconductor does not lead to a strong suppression of screening by charge when $V_\text{SC}=0$: the field effect strongly reduces as soon as charge enters the wire when we take charge screening into account.
This lack of interplay between the screening by superconductor and by charge can be understood by looking at the charge density distribution in the nanowire [Fig.~\ref{fig:screening}(b), (c)].
Since a positive gate voltage is required to induce a finite charge density, the charges are pulled away from the superconductor, and the corresponding mirror charges in the superconductor area located at a distance comparable to twice the wire thickness.
On the contrary, a positive $V_\textrm{SC}$ requires a compensating negative $V_\textrm{G}$ to induce comparable charge density in the wire, pushing the charges closer to the superconductor [Fig.~\ref{fig:screening}(d)].
In this case, the proximity of the electron density to the superconductor leads to the largest suppression of the lever arm, and proximity of image charges almost completely compensates the screening by charge.

The Van Hove singularity in the density of states leads to an observable kink in $\mu_i$ each time an extra band crosses the Fermi level [inset in Fig.~\ref{fig:energy_levels}(a)].
However, we observe that the effect is weak on the scale of level spacing and cannot guarantee strong pinning of the Fermi level to a band bottom.

\section{Electrostatic response to the Zeeman field}
\label{sec:magresponse}

\subsection{Limit of large level spacing}

\co{Motivation to use perturbation theory instead of full self-consistency}

The full self-consistent solution of the Schr\"odinger-Poisson equation is computationally expensive and also hard to interpret due to a high dimensionality of the space of unknown variables.
We find a simpler form of the solution at a finite Zeeman field relying on the large level spacing $\sim$\SI{10}{\meV} in typical nanowires.
It ensures that the transverse wave functions stay approximately constant, i.e. $|\langle \psi(E_{\textrm{Z}}) | \psi(0) \rangle | \approx 1 $ up to magnetic fields of $\sim \SI{7}{\tesla}$.
In this limit we may apply perturbation theory to compute corrections to the chemical potential for varying $E_{\textrm{Z}}$.

\co{Derivation of perturbation in mu due to Ez}

We write the potential distribution for a given $E_\textrm{z}$ in the form 
\begin{equation}
\phi(x, y, E_\text{Z}) = \phi_{\textrm{b.c.}}(x,y) + \sum_{i=0}^N \phi_i(x,y,E_\textrm{z}),
\label{eq:initphi}
\end{equation}
where $\phi_{\textrm{b.c.}}$ is the potential obeying the boundary conditions set by the gate and the superconducting lead, and solves the Laplace equation
\begin{equation}
\nabla^2 \phi_{\textrm{b.c.}}(x,y) = 0.
\label{eq:initLaplace}
\end{equation}
The corrections $\phi_i$ to this potential due to the charge contributed by the
$i$-th mode out of the $N$ modes below the Fermi level then obeys a 
Poisson equation with Dirichlet boundary conditions (zero voltage
on the gates):
\begin{equation}
\nabla^2 \phi_i(x,y,E_\text{Z}) = \frac{e}{\epsilon} \left|\psi_i(x,y)\right|^2
n(-\mu_i-\delta \mu_i, E_\text{Z}, \alpha) 
\end{equation}
where we write the chemical potential at a finite value of $E_\text{Z}$ as
$\mu_i(E_z) = \mu_i +\delta \mu_i$ where $\mu_i$ is the chemical potential in the
absence of a field.

We now define a magnetic field-independent reciprocal capacitance as
\begin{equation}
P_i(x,y) = \frac{\phi_i(x,y, E_\text{Z})}{-e\, n(-\mu_i-\delta \mu_i, E_\text{Z}, \alpha)} \label{eq:reciprocal_capacitance}
\end{equation}
which solves the Poisson equation
\begin{equation}
\nabla^2 P_i(x, y) = -\frac{1}{\epsilon} |\psi_i(x,y)|^2\,.
\end{equation}

Having solved the Schr\"odinger-Poisson problem numerically for
$E_\text{Z}=0$, we define $\delta \phi_i = \phi_i(x,y, E_\text{Z})
-\phi_i(x, y, 0)$ and $\delta n = n(-\mu_i-\delta \mu_i, E_\text{Z}, \alpha)
-n(-\mu_i, 0, \alpha)$. The correction $\delta E_i$ to the subband energy $E_i$ is then given in first order perturbation as
\begin{equation}
\delta E_i = -e \langle \psi_i| \sum_{j=0}^N \delta \phi_j | \psi_i \rangle\,.
\label{eq:firstordershift}
\end{equation}
Using Eqs.~\eqref{eq:reciprocal_capacitance}, \eqref{eq:firstordershift} and $\delta \mu_i = - \delta E_i$ we then arrive at:
\begin{equation}
\delta \mu_i = -e^2 \sum_{j=0}^N P_{ij} \delta n_j\,,
\label{eq:deltamufinal}
\end{equation}
with the elements of the reciprocal capacitance matrix $P$ given by
\begin{equation}
P_{ij} = \langle \psi_i | P_j  | \psi_i \rangle.
\label{eq:capacity_elements}
\end{equation}
Solving the Eq.~\eqref{eq:deltamufinal} self-consistently, we compute corrections to the initial chemical potentials $\mu_i$.
The Eq.~\eqref{eq:deltamufinal} has a much lower dimensionality than Eq.~\eqref{eq:minimize_func} and is much cheaper to solve numerically.
Further, all the electrostatic phenomena enter Eq.~\eqref{eq:deltamufinal} only through the reciprocal capacitance matrix Eq. \eqref{eq:capacity_elements}.

\subsection{Single- and multiband response to the magnetic field}

\co{Response of mu to Ez: screening and spin-orbit reduce tunability}

We start by computing the electrostatic response to changes in the magnetic field when the Fermi level is close to the band bottom for a single band ($N=1$, and we write the index $\mu_1 \equiv \mu$ for brevity).
We study the influence of the electrostatic environment and assess whether the device is closer to a constant charge density or constant chemical potential situation (using the nomenclature of Ref.~\onlinecite{Sarma2012} explained in App.~\ref{app:nomenclature}).

\begin{figure}[tb]
\centerline{\includegraphics[width=\linewidth, trim = 0 3cm 0 2.3cm]{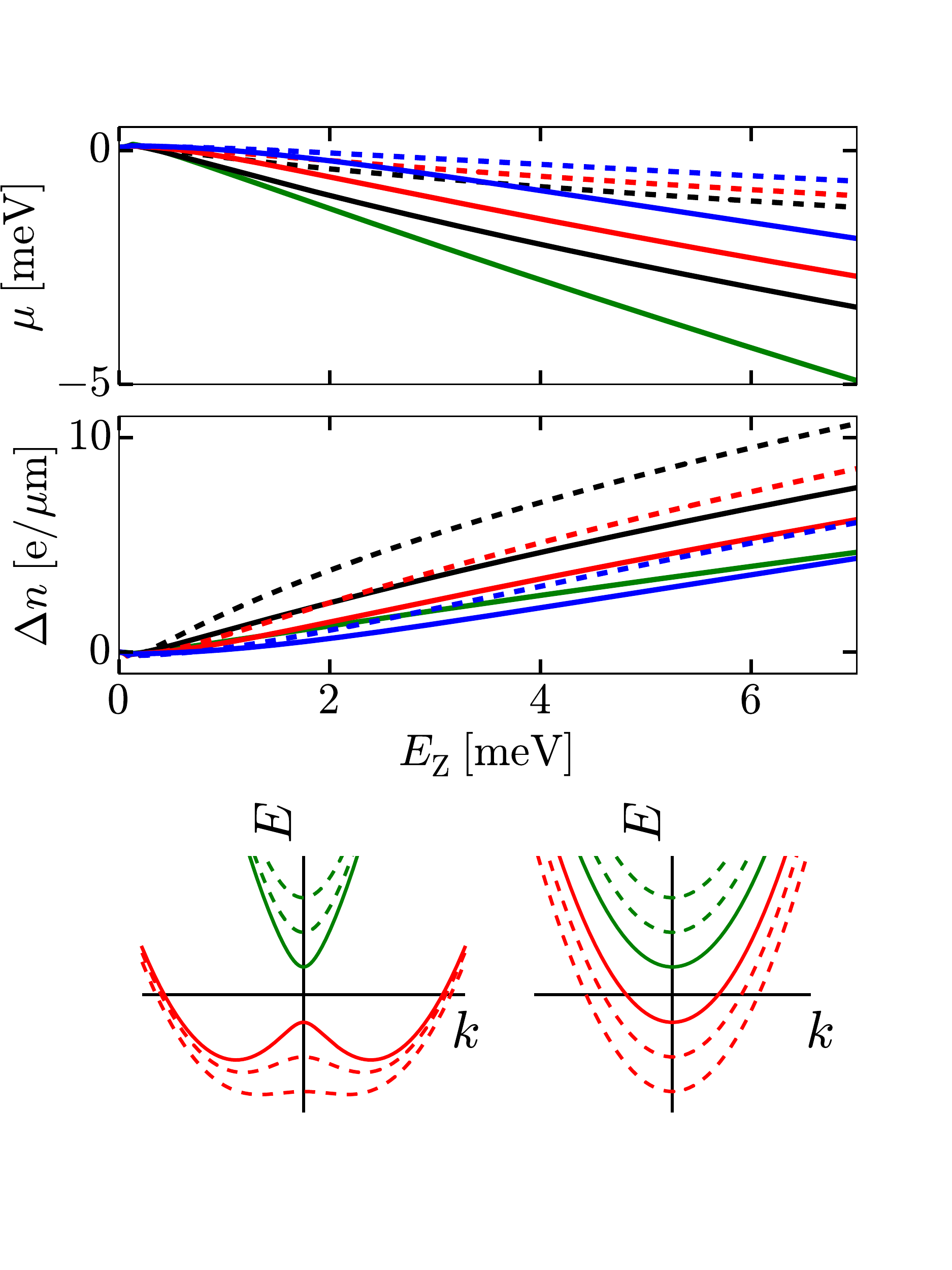}}
\caption{Top and middle panel:
Variation in chemical potential (top panel) and in electron density (middle panel) as a function of magnetic field.
The green solid line corresponds to the case without a superconductor.
Other solid lines correspond to $V_{\textrm{SC}} = \SI{0}{\volt}$, dashed lines to $V_{\textrm{SC}} = \SI{0.2}{\volt}$.
Black, red and blue indicate spin-orbit lengths of 233, 100, and \SI{60}{\nano\meter} respectively. Bottom panel: Dispersion relation $E(k)$ for $E_\text{SO} \gg E_\text{Z}$ (left) and $E_\text{SO} \ll E_\text{Z}$ (right). Dashed lines indicate the evolution of the dispersion for the increasing magnetic field.}
\label{fig:mu_variation}
\end{figure}

The top panel of Fig.~\ref{fig:mu_variation} shows the chemical potential response to Zeeman field.
Without a superconducting contact, the electron-electron interactions in the nanowire are screened the least, and the Coulomb effects are the strongest, counteracting density changes in the wire.
In agreement with this observation, we find the change in chemical potential $\mu$ comparable to the change in $E_{\textrm{Z}}$.
Hence, in this case the system is close to a constant-density regime.

A superconducting contact close to the nanowire screens the electron-electron interaction in the wire due to image charges.
The chemical potential is then less sensitive to changes in magnetic field.
We find that this effect is most pronounced for a positive work function difference with the superconductor $V_{\textrm{SC}} = \SI{0.2}{\volt}$, when most of the electrons are pulled close to the superconducting contact.
Then, the image charges are close to the electrons and strongly reduce the Coulomb interactions.
In this case the system is close to a constant chemical potential regime.
For $V_\text{SC} = \SI{0}{\volt}$ screening from the superconducting contact is less effective, since electric charges are further away from the interface with the superconductor.
Therefore in this case, we find a behavior intermediate between constant density and constant chemical potential.

Besides the dependence on the electrostatic surrounding, the magnetic field response of the chemical potential depends on the spin-orbit strength.
Specifically, the chemical potential stays constant over a longer field range when the spin-orbit interaction is stronger.
\footnote{Although we decrease the spin-orbit length to $l_\text{SO} = \SI{60}{\nano\meter}$, which is smaller than the wire diameter of \SI{100}{\nano\meter}, we assume separable wave functions.
Screening by the superconductor strongly localizes the wave functions, such that the confinement is still smaller than the spin-orbit length.}
The bottom panel of Fig.~\ref{fig:mu_variation} explains this: when the spin-orbit energy $E_\text{SO} \gg E_\text{Z}$, the lower band has a W-shape (bottom left). A magnetic-field increase initially transforms the lower band back from a W-shape to a parabolic band, as indicated by the dashed red lines.
During this transition, the Fermi wavelength is almost constant.
Since the electron density is proportional to the Fermi wavelength, this means that both the density and the chemical potential change very little in this regime.
We thus identify the spin-orbit interaction as another phenomenon driving the system closer to the constant chemical potential regime, similar to the screening of the Coulomb interaction by the superconductor.

At large Zeeman energies $E_{\textrm{Z}} \gtrsim E_\text{SO}$, the spin-down band becomes parabolic (bottom right of Fig.~\ref{fig:mu_variation}).
This results in the slope of $\mu(E_\text{Z})$ becoming independent of the spin-orbit coupling strength, as seen in the top panel of Fig.~\ref{fig:mu_variation} at large values of $E_\text{Z}$.

\co{Limits of \mu and n for E_Z << P^2 and E_Z >> P^2}
Close to the band bottom and when spin-orbit interaction is negligible, we study the asymptotic behavior of $\mu$ and $n$ by combining the appropriate density expression Eq.~\ref{eq:middledensity} with the corrections in the chemical potential Eq.~\ref{eq:deltamufinal}.
In that case, the chemical potential becomes
\begin{equation}
\mu = -\frac{e^2 P}{\pi \hbar} \sqrt{2m^*(\mu + E_\text{Z})}.
\label{eq:mu_proportionality}
\end{equation}
We associate an energy scale $E_P$ with the reciprocal capacitance $P$, given by 
\begin{equation}
E_P = \frac{2m^* e^4 P^2}{\pi^2 \hbar^2},
\label{eq:capacitance_energyscale}
\end{equation}
and study the two limits $E_P \gg E_\text{Z}$ and $E_P \ll E_\text{Z}$.
In the strong screening limit $E_P \gg E_\text{Z}$ we find the asymptotic behavior $\mu \approx -E_\text{Z}$, corresponding to a constant-density regime. The opposite limit $E_P \ll E_\text{Z}$ yields $\mu \approx -\sqrt{E_P E_\text{Z}}$, close to a constant chemical potential regime. We computed $E_P$ explicitly for the chemical potential variations as shown in the top panel of Fig.~\ref{fig:mu_variation}. For a nanowire without a superconducting lead, we find an energy $E_P \approx \SI{42}{\milli\eV} \gg E_\text{Z}$, indicating a constant-density regime. Using the classical approximation of a metallic cylinder above a metallic plate, we find an energy of the same order of magnitude. For a nanowire with an attached superconducting lead at $V_\text{SC} = \SI{0}{\volt}$, we get $E_P \approx \SI{7}{\meV} \sim E_\text{Z}$, intermediate between constant density and constant chemical potential. Finally, a superconducting lead at $V_\text{SC} = \SI{0.2}{\volt}$ yields $E_P \approx \SI{0.5}{\meV} \ll E_\text{Z}$, indicating a system close to the constant chemical potential regime.

\co{Reciprocal capacitance is measurable.}
Since integrating over density-of-states measurements yields $\delta n_0$, the inverse self-capacitance $ -e \langle \psi_0 | P_0 | \psi_0 \rangle$ can be inferred from experimental data by fitting the density variation curves to the theoretical dependence $\mu(E_\text{Z})$.
This allows to experimentally measure the effect of the electrostatic environment, knowing the remaining Hamiltonian parameters.

\begin{figure}[tbh]
\centerline{\includegraphics[width=.9\linewidth]{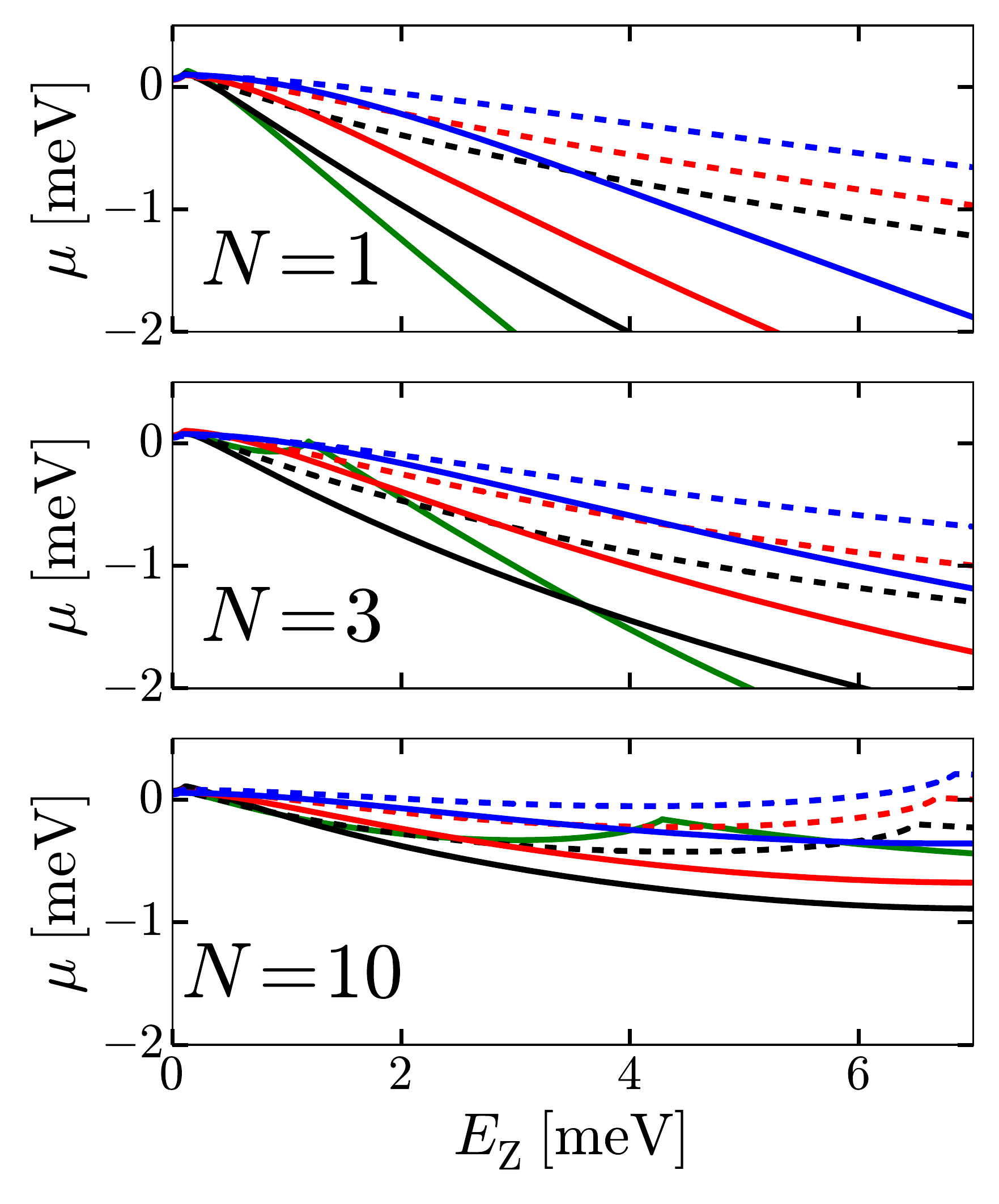}}
\caption{Response of $\mu_N$ as a function of magnetic field for $N=1,3$, and $10$, all close to the band bottom.
The solid green line corresponds to the case of no screening by a superconductor.
Other solid lines correspond to $V_{\textrm{SC}} = \SI{0}{\volt}$, dashed lines to $V_{\textrm{SC}} = \SI{0.2}{\volt}$.
Black, red and blue indicate spin-orbit lengths of 233, 100, and \SI{60}{\nano\meter} respectively.}
\label{fig:mu_multiple}
\end{figure}

\co{Multiple-mode response to Ez; effects and control of mu are less for more modes}
We compare the response to Zeeman field in the multi-band case for $N=3$ and $N=10$ to the single band behavior in Fig.~\ref{fig:mu_multiple}.
We observe that presence of extra charges further reduces the sensitivity of the chemical potential to the magnetic field.
We interpret the non-monotonous behavior of the chemical potential (most pronounced for $N=10$ in Fig.~\ref{fig:mu_multiple}, but in principle present for all $N$) as being due to a combination of the Van Hove singularities in the density of states and screening by charges.
For a fixed chemical potential, the upper band, moving up in energy due to the magnetic field, loses more states than the lower band acquires, since it approaches the Van Hove singularity in its density of states.
To keep the overall density fixed, the chemical potential increases.
Once the density in the lower band equals the initial density, the upper band is empty and the chemical potential starts dropping again. In the limit of
constant density and a single mode the magnetic field dependence of the
chemical potential can be solved analytically, reproducing the
non-monotonicity and kinks (see App.~\ref{sec:constantdens_analytics}).

Relating the variation in $\mu_i$ to density measurements is experimentally inaccessible for $N > 1$, since corrections to $\mu_i$ depend on the density changes of each individual mode, as expressed in Eq.~\eqref{eq:deltamufinal}.

\section{Impact of electrostatics on Majorana properties}
\label{sec:majoprop}

\subsection{Shape of the Majorana phase boundary}

\co{The phase transition depends on mu and thus on electrostatics}

The nanowire enters the topological phase when the bulk energy gap closes at a Zeeman energy of $E_{\textrm{Z}} = \sqrt{\mu^2 + \Delta^2}$.
The electrostatic effects affect the shape of the topological phase boundary through the dependence of $\mu$ on $E_\text{Z}$.
To find the topological phase boundary as a function of both experimentally controllable parameters $V_{\textrm{G}}$ and $E_{\textrm{Z}}$, we perform a full self-consistent simulation at $E_{\textrm{Z}} = 0$.
We then compute corrections to the resulting chemical potential at arbitrary $E_{\textrm{Z}}$ using Eq.~\eqref{eq:deltamufinal}, and find topological phase boundary $E_{\textrm{Z}} = \sqrt{\mu^2 + \Delta^2}$ by recursive bisection.

\co{Resulting phase boundary}

\begin{figure}[tb]
\centerline{\includegraphics[width=1.0 \linewidth]{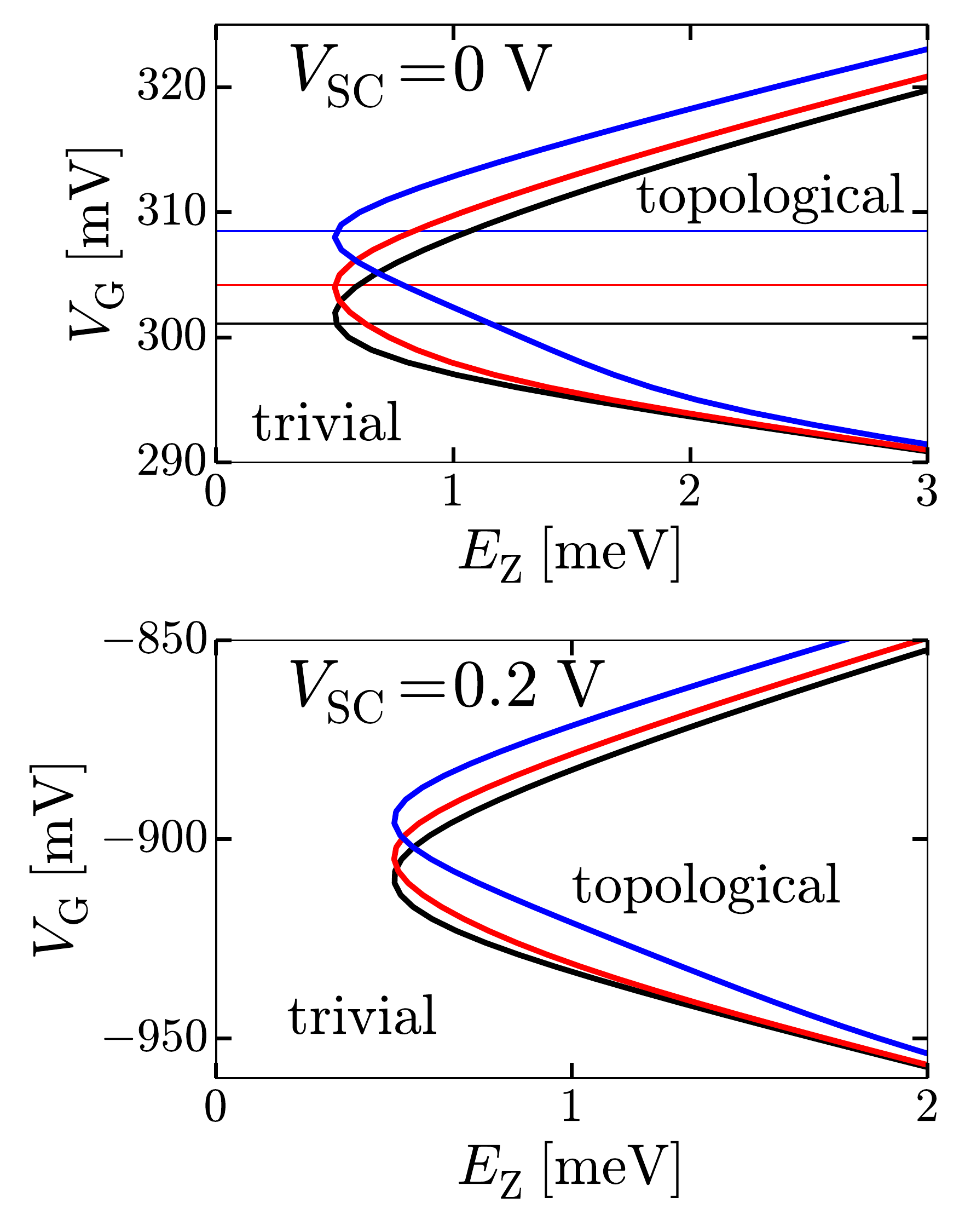}}
\caption{Majorana transition boundary for a superconductor at $V_{\textrm{SC}} = \SI{0}{\volt}$ (upper panel) or a superconductor at $V_{\textrm{SC}} = \SI{0.2}{\volt}$ (lower panel).
The superconducting gap $\Delta = \SI{0.5}{\meV}$.
The boundaries are obtained for the single-band case.
The solid black, red, and blue lines correspond to a spin-orbit length of 233, 100, and \SI{60}{\nm} respectively.
The black, red and blue horizontal lines in the upper plot indicate the gate voltages at which we compute the correspondingly colored Majorana coupling oscillations in the inset of Fig.~\ref{fig:smoking_gun}.}
\label{fig:phase_bound}
\end{figure}

Figure~\ref{fig:phase_bound} shows the resulting phase boundary corresponding to $\Delta = \SI{0.5}{\meV}$.
The phase boundary has a non-universal shape due to the interplay between electrostatics and magnetic field. 
In agreement with our previous conclusions, the electrostatic effects are the
strongest with absent work function difference $V_{\textrm{SC}} = \SI{0}{\volt}$ (top panel of Fig.~\ref{fig:phase_bound}) when the nanowire is intermediate between constant density and constant chemical potential. \footnote{The presence of a superconductor is essential for Majorana fermions, but inevitably leads to screening. For the geometries of our calculations we thus do not have a situation close to constant density.}
Close to the band bottom, the charge screening reduces changes in density, and
thus lowers the chemical potential by an amount that is similar to $E_\text{Z}$. Hence, the lower phase boundary (at smaller $V_\text{G}$) has a weaker slope than the upper phase boundary (at larger $V_\text{G}$). Note that in the limit of constant density, the lower phase boundary would be a constant independent of $E_\text{Z}$ (see App.~\ref{sec:constantdens_analytics}). 

For a work function difference $V_{\textrm{SC}} = \SI{0.2}{\volt}$, the system
is closer to the constant chemical potential regime.
In this regime, $\mu$ changes linearly with $V_\text{G}$, yielding a hyperbolic phase boundary with symmetric upper and lower arms and its vertex at $E_\text{Z} = \Delta$.
When spin-orbit interaction is strong, a transition in the lower arm of the phase boundary from constant chemical potential (hyperbolic phase boundary) to constant density (more horizontal lower arm) occurs, resulting in a `wiggle' which is most pronounced for $V_{\textrm{SC}} = \SI{0}{\volt}$ and $l_\text{SO} = \SI{60}{\nm}$.
This feature is less pronounced for $V_{\textrm{SC}} = \SI{0.2}{\volt}$ due to the screening by the superconductor suppressing the Coulomb interactions.

\subsection{Oscillations of Majorana coupling energy}

\co{Majorana energy oscillations in finite-length nanowires provide `smoking gun' evidence for their existence}

The wave functions of the two Majorana modes at the endpoints of a finite-length nanowire have a finite overlap that results in a finite nonzero energy splitting $\Delta E$ of the lowest Hamiltonian eigenstates~\cite{Cheng2009, Cheng2010, Prada2012, Sarma2012, Rainis2013}.
This splitting oscillates as a function of the effective Fermi wave vector $k_{\textrm{F,eff}}$ as $\textrm{cos}(k_{\textrm{F,eff}} L)$
\cite{Sarma2012}.
We investigate the dependency of the oscillation frequency, or the oscillation peak spacing on magnetic field and the electrostatic environment.

\co{peak spacing: analytic expression}

A peak in the Majorana splitting energy occurs when Majorana wave functions
constructively interfere, or when the Fermi momentum equals $q \pi /L$, with $q$ the peak number and $L$ the nanowire length.
The momentum difference between two peaks is
\begin{equation}
k_{\textrm{F,eff}}(E_{\textrm{Z}, q+1}) - k_{\textrm{F,eff}}(E_{\textrm{Z},q}) = \frac{\pi}{L},
\label{eq:momentum_distance}
\end{equation}
where $E_{\textrm{Z},q}$ is the Zeeman energy corresponding to the $q$-th oscillation peak.
In the limit of small peak spacing, we expand $k_{\textrm{F,eff}}(E_{\textrm{Z}, q+1}) - k_{\textrm{F,eff}}(E_{\textrm{Z},q})$ to first order in $E_\text{Z}$:
\begin{equation}
\frac{dk_{\textrm{F}}}{dE_{\textrm{Z}}} \Delta E_{\textrm{Z}} = \frac{\pi}{L},
\end{equation}
yielding the peak spacing
\begin{equation}
\Delta E_{\textrm{Z, peak}} = \frac{\pi}{L} \left( \frac{dk_{\textrm{F}}}{dE_{\textrm{Z}}} \right) ^{-1}.
\end{equation}
Since $k_{\textrm{F,eff}} = k_{\textrm{F,eff}}(E_{\textrm{Z}}, \mu(E_{\textrm{Z}}))$, we substitute
\begin{equation}
\frac{dk_{\textrm{F}}}{dE_{\textrm{Z}}} = \frac{\partial k_{\textrm{F}}}{\partial E_{\textrm{Z}}} + \frac{\partial k_{\textrm{F}}}{\partial \mu}\frac{d\mu}{dE_{\textrm{Z}}}.
\end{equation}
We obtain the values of $\partial k_{\textrm{F}} / \partial E_{\textrm{Z}}$ and $\partial k_{\textrm{F}} / \partial \mu$ from the analytic expression for $k_{\textrm{F}}$, presented in App.~\ref{app:density}.
The value $d\mu / dE_{\textrm{Z}}$ results from the dependence $\mu(E_{\textrm{Z}})$ shown in Fig.~\ref{fig:mu_variation}.

\begin{figure}[tb]
\centerline{\includegraphics[width=1.0\linewidth]{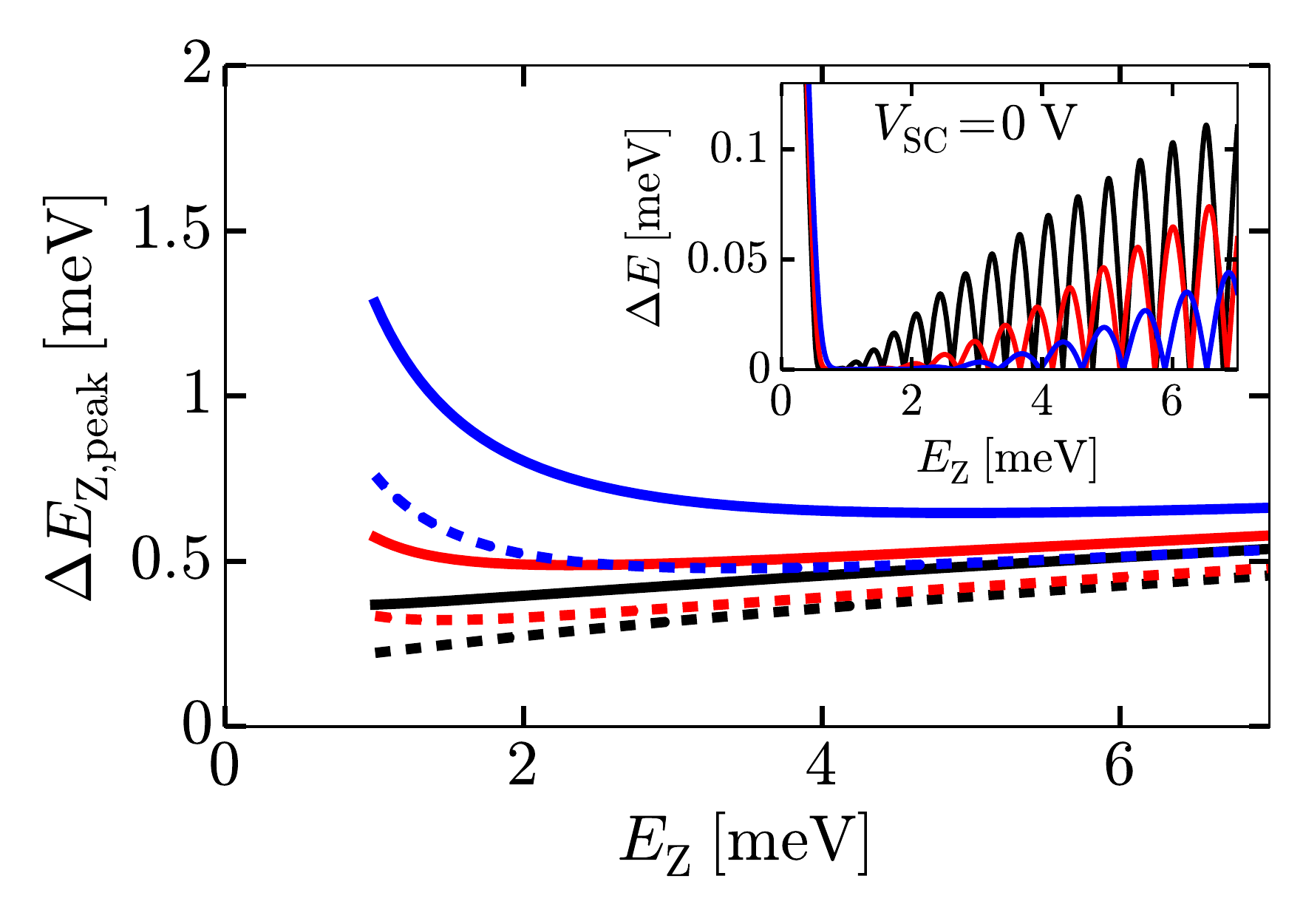}}
\caption{Peak spacing of the Majorana energy oscillations in a magnetic field  for a nanowire of length $L=\SI{2}{\um}$.
Solid lines correspond to $V_{\textrm{SC}} = \SI{0}{\volt}$, dashed lines to $V_{\textrm{SC}} = \SI{0.2}{\volt}$.
Black, red and blue indicate spin-orbit lengths of 233, 100, and \SI{60}{\nm} respectively.
Inset: Splitting energy oscillations for $V_{\textrm{SC}} = \SI{0}{\volt}$.
The three horizontal lines in the upper panel of Fig.~\ref{fig:phase_bound} indicate the corresponding gate potential. The energy splittings are found by solving for the lowest energy of the Hamiltonian of Eq.~\eqref{eq:BdGHamiltonian1D}, using the chemical potentials obtained from the perturbation scheme as described in Sec.~\ref{sec:magresponse}.}
\label{fig:smoking_gun}
\end{figure}

\co{This has consequences for the characteristics and the supposed universality of the peak spacing}

Fig.~\ref{fig:smoking_gun} shows the peak spacing as a function of $E_{\textrm{Z}}$ for a nanowire of length $L=\SI{2}{\um}$.
Stronger screening reduces the peak spacing (i.e. increases the oscillation frequency) by reducing the sensitivity of the chemical potential to the magnetic field, as discussed in Sec.~\ref{sec:magresponse}.
In addition, spin-orbit strength has a strong influence on the peak spacing, since for $E_Z \ll E_\text{SO}$ the density and thus $k_\text{F,eff}$ stay constant.
This results in a lower oscillation frequency and hence a larger peak spacing. Correspondingly, we find that the peak spacing may increase, decrease, or roughly stay constant as a function of the magnetic field.

Similarly to the shape of the Majorana transition boundary, Fig.~\ref{fig:smoking_gun} shows that the peak  spacing does not follow a universal law, in contrast to earlier predictions \cite{Rainis2013}.
In particular, our findings may explain the zero-bias oscillations measured in Ref.~\onlinecite{Churchill2013}, exhibiting a roughly constant peak spacing.
\begin{figure}[tb]
\centerline{\includegraphics[width=1.0\linewidth]{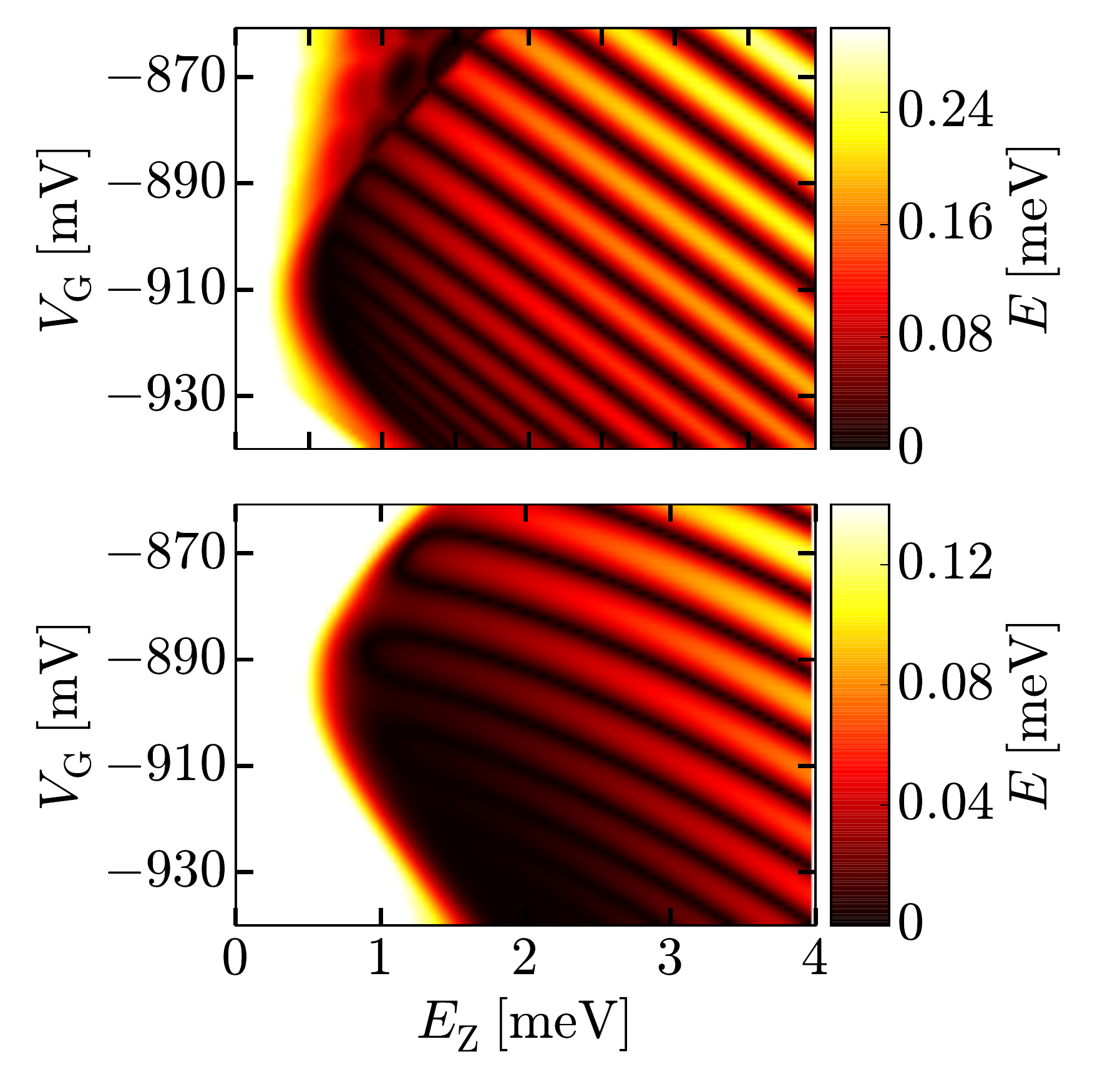}}
\caption{Majorana energy oscillations as a function of gate voltage and magnetic field for a superconductor at $V_\text{SC} = \SI{0.2}{\volt}$ with weak spin-orbit interaction, $l_\text{SO} = \SI{233}{\nm}$ (upper panel), and strong spin-orbit interaction, $l_\text{SO} = \SI{60}{\nm}$ (lower panel).}
\label{fig:majo_osc_space}
\end{figure}

\co{Coupling oscillations in V_G - E_Z space}

Fig.~\ref{fig:majo_osc_space} shows Majorana energy oscillations as a function of both gate voltage and magnetic field strength for $V_\text{SC}=\SI{0.2}{\volt}$, with $L = \SI{1000}{\nm}$ to increase the Majorana coupling.
The diagonal ridges are lines of constant chemical potential.
The difference in slope between the ridges of both plots indicates a difference in the equilibrium situation: closer to constant density for weak spin-orbit coupling, closer to constant chemical potential for strong spin-orbit coupling.
The bending of the constant chemical potential lines in the lower panel indicates a transition from the latter mechanism to the former mechanism, due to the increase of magnetic field, as explained in Sec.~\ref{sec:magresponse}.

\section{Summary}
\label{sec:conclusions}

\co{Screening reduces gate field effect}

We have studied the effects of the electrostatic environment on the field control of Majorana devices and their properties.
Screening by charge and by the superconductor strongly reduce the field effect of the gates.
Furthermore, screening by the superconductor localizes the charge and induces a large internal electric field.
When we assume the superconductor to have a zero work function difference with the nanowire, charge localizes at the bottom of the wire, which reduces the induced superconducting gap.

\co{Screening and spin-orbit take the system away from the constant-density regime}

Coulomb interaction causes the chemical potential to respond to an applied magnetic field, while screening by the superconductor and spin-orbit interaction suppress this effect.
If a superconductor is attached, the equilibrium regime is no longer close to constant density, but either intermediate between constant density and constant chemical potential for a superconductor with zero work function difference, or close to constant chemical potential for a superconductor with a positive work function difference. An increasing spin-orbit interaction also reduces the response of the chemical potential.

\co{Due to their dependency on mu, the two Majorana signatures are sensitive to the electrostatic environment}

Due to this transition in equilibrium regime for increasing screening and spin-orbit interaction, the shape of the Majorana phase boundary and the oscillations of Majorana splitting energy, depend on device parameters instead of following a universal law.

\co{If parameter can be measured experimentally, we can compare direct measurements of Majorana signatures with the predictions derived from this parameter. }

We have shown how to relate the measurement of density variations to the chemical potential response.
Since the Majorana signatures directly depend on this response, our work offers a way to compare direct experimental observations of both signatures with theoretical predictions, and to remove the uncertainty caused by the electrostatic environment.

\co{Finally, SP simulations can be used to compute lever arms and capacity in nanowire devices}

Our Schr\"odinger-Poison solver, available in the supplementary files for this manuscript, can be used to compute lever arms and capacities for different device dimensions and geometries, providing practical help for the design and control of experimental devices.

\acknowledgments

We thank R. J. Skolasiński for reviewing the code, T. Hyart and P. Benedysiuk for valuable discussion.
This research was supported by the Foundation for Fundamental Research on Matter (FOM), Microsoft Corporation Station Q, the Netherlands Organization for Scientific Research (NWO/OCW) as part of the Frontiers of Nanoscience program, and an ERC Starting Grant.

\appendix

\section{Nomenclature -- constant density and constant chemical potential}
\label{app:nomenclature}

\begin{figure}
\includegraphics[width=\linewidth]{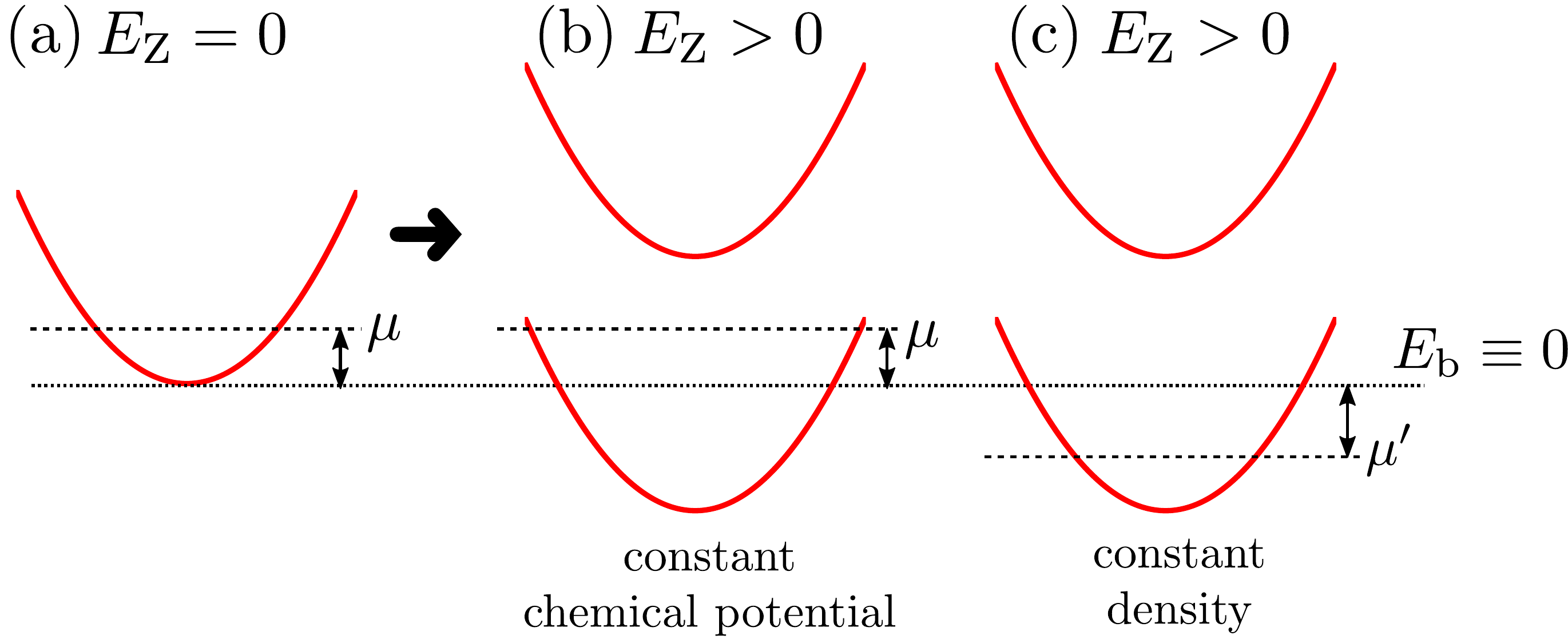}
\caption{Schematic explanation of constant chemical potential and
  constant density limits discussed in
  Ref.~\onlinecite{Sarma2012}: (a) In the absence of a magnetic field,
  a band is filled up to the chemical potential $\mu$. $\mu$ is
  measured with respect to the band edge $E_\text{b}$ that serves as a
  reference energy. For a finite Zeeman splitting $E_\text{Z}$ the two
  spin-bands split by $\pm E_\text{Z}$ with respect to
  $E_\text{b}$. In this case there can be two extreme situations: (b)
  constant chemical potential -- $\mu$ stays unchanged (and hence
  the total electron density changes). (c) constant density -- the
  total electron density stays constant leading to a new chemical
  potential $\mu'$. (For simplicity, all plots are shown for
  $\alpha=0$.)}
\label{fig:const_vs_const}
\end{figure}

In Ref.~\onlinecite{Sarma2012} Das Sarma \emph{et al.}\ considered
Majorana oscillations as a function of magnetic field. The authors
considered there two extreme electrostatic situations that they refer to as
constant chemical potential and constant density.

In particular, Ref.~\onlinecite{Sarma2012} considers a one-dimensional
nanowire BdG Hamiltonian as in Eq.~\eqref{eq:BdGHamiltonian1D}, with
$\mu_1$ being denoted as $\mu$. In this model, the subband energy
$E_\text{b}$ is fixed and set to $0$. The electron density is changed by
adjusting $\mu$ (shown for the $E_\text{Z}=0$ case in
Fig.~\ref{fig:const_vs_const}(a).

For fixed $\mu$ in Eq.~\eqref{eq:BdGHamiltonian1D} electron density will change
upon changing $E_\text{Z}$. For example, if $E_\text{Z} \gg \mu, E_\text{so}$,
electron density will increase monotonically as $E_\text{Z}$ is increased
(see Fig.~\ref{fig:const_vs_const}(b). This constant chemical potential
situation is realized in the limit of vanishing Coulomb
interaction, as then density changes do not influence the electrostatic
potential. The same assumption is used in Refs.~\onlinecite{Prada2012,
Rainis2013}.

Ref.~\onlinecite{Sarma2012} also considered the opposite case
of infinitely strong Coulomb interaction. In this case the electron density
is fixed, and consequently $\mu$ must change as $E_\text{Z}$ changes.
This constant density situation is schematically shown in
Fig.~\ref{fig:const_vs_const}(c).

\section{Lever arms in an InAs-Al nanowire}
\label{app:marcus_device}

\co{Introduce the Marcus device}
Another promising set of devices for the creation of Majorana zero modes is an epitaxial InAs-Al semiconductor–superconductor nanowire.
These systems exhibit a hard superconducting gap and a high interface quality due to the epitaxial growth of the Al superconductor shell \cite{Chang2014}.

\co{Explain the device and its boundary conditions}

\begin{figure}[tb]
\centerline{\includegraphics[width=1.0\linewidth]{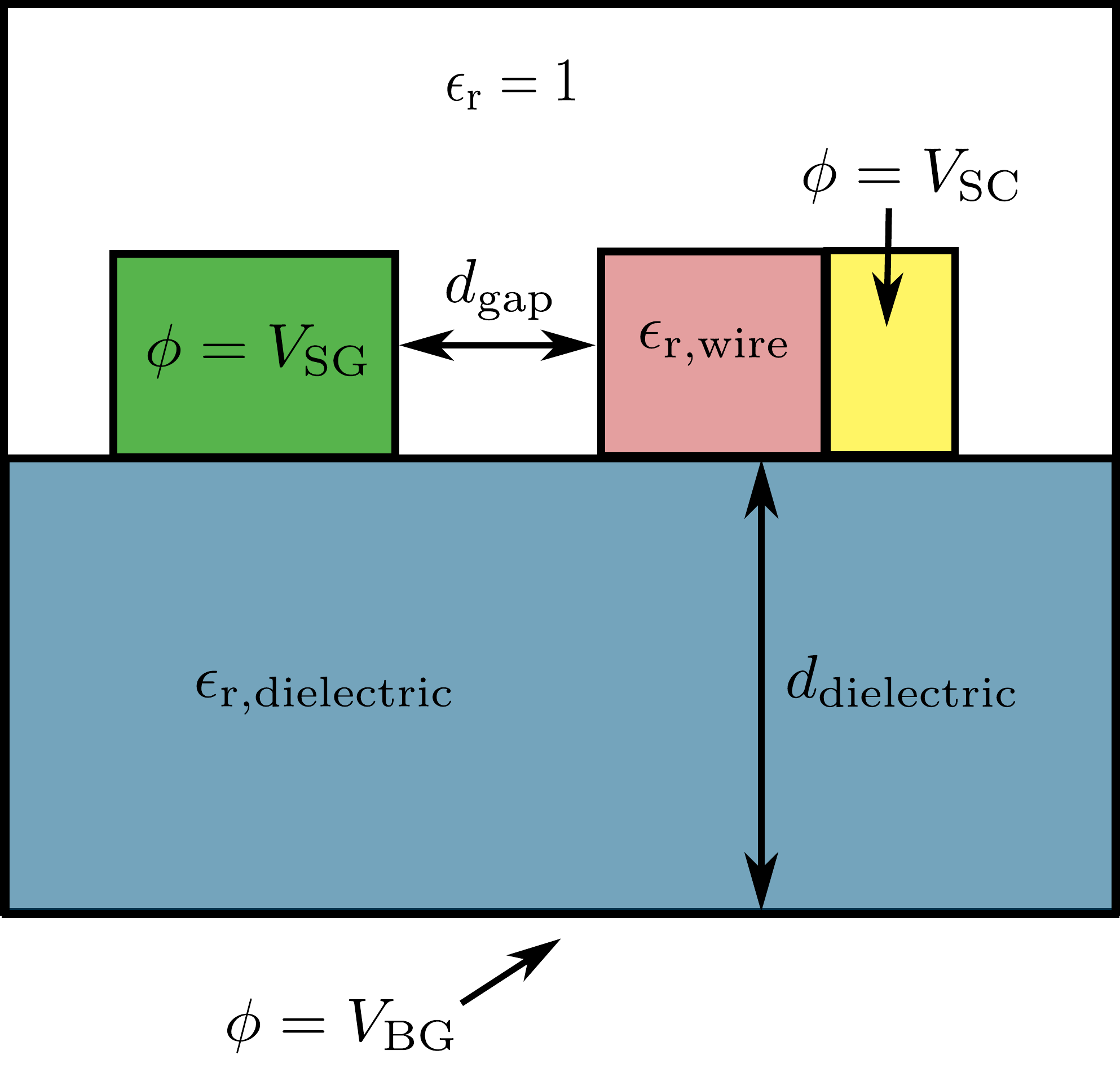}}
\caption{Schematic picture of the cross section of an InAs-Al device.
It consists of a nanowire with a square cross section on a  dielectric layer which covers a global back gate.
A superconducting lead covers one side of the wire.
A vacuum gap separates the wire from a second gate.}
\label{fig:marcus_device}
\end{figure}

Figure~\ref{fig:marcus_device} shows a cross section of the device.
The $\epsilon_\text{r} = 14.6$ nanowire (InAs) lies on an  $\epsilon_\text{r} = 4$ dielectric layer (SiO$_2$) of thickness $d_{\textrm{dielectric}}=\SI{200}{\nm}$ and is connected on one side to an Al superconducting shell.
The device has two gates: a global back-gate with a gate potential $V_{\textrm{BG}}$, and a side gate with a potential $V_{\textrm{SG}}$, separated by a vacuum gap of width $d_{\textrm{gap}}$.
We model the superconductor again as a metal with a fixed potential $V_{\textrm{SC}}$.
These three potentials form the boundary conditions of the system.

\co{Explain method: Schrodinger-Poisson}

We estimate the dependence of the lever arm of the side date $dE/d V_\textrm{SG}$ on $d_{\textrm{gap}}$ using the self-consistent Schr\"odinger-Poisson simulations.
We set the back gate to $V_{\textrm{BG}} =\SI{-3.5}{\volt}$, and choose the work function difference of the Al shell equal to \SI{0.26}{\eV}, such that one electron mode is present at a side gate voltage of $V_{\textrm{SG}} = \SI{-2}{\volt}$, with $d_{\textrm{gap}} = \SI{145}{\nm}$, as was observed in experiments~\cite{Deng2015}. We use the band gap \SI{0.36}{\eV} for InAs.

\begin{figure}[tb]
\centerline{\includegraphics[width=1.0 \linewidth]{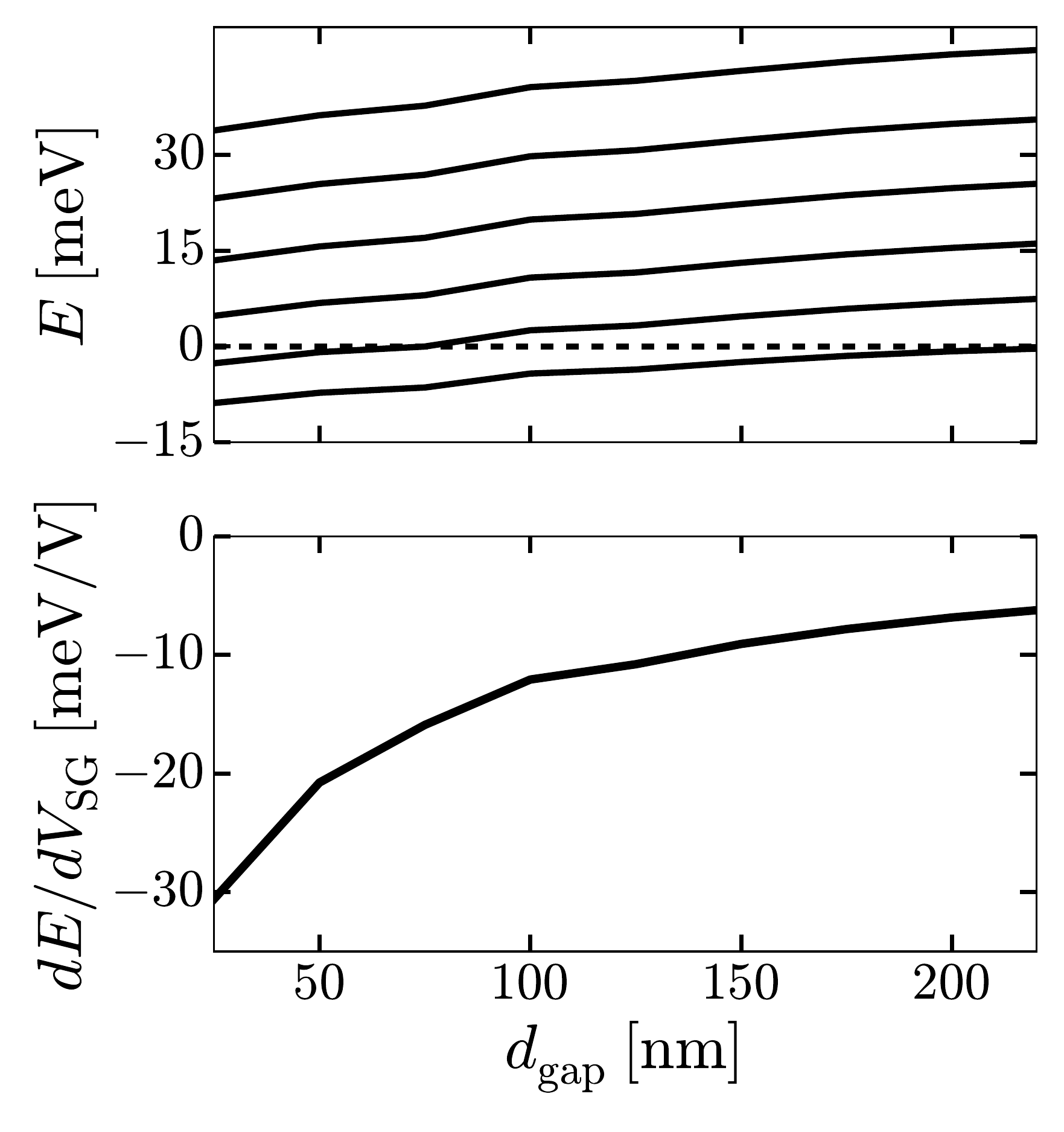}}
\caption{Top panel: six lowest energy levels with a fixed gate potential $V_{\textrm{BG}}=V_{\textrm{SG}}=\SI{0}{\volt}$.
Bottom panel: lever arm in the InAs-Al device as a function of gate spacing with $V_{\textrm{BG}}=\SI{0}{\volt}$.}
\label{fig:gap_leverarm}
\end{figure}

\co{Message: self-consistent SP simulations can help experimentalists solving practical issues}

Our results are shown in Fig.~\ref{fig:gap_leverarm}, and allow to translate the gate voltages into the nanowire chemical potential.
The work for the InAs-Al device shows that our numerical algorithm is easily adjusted to different device geometries, as long as the nanowire stays translationally invariant.

\section{Electron density in a nanowire}
\label{app:density}

Integration over the 1D density of states yields the electron density $n(E, E_Z, \alpha)$, related to the charge density by Eq.~\eqref{eq:density}.
To derive the density of states, we start from the nanowire Hamiltonian, consisting of the transverse Hamiltonian of Eq.~\eqref{eq:transverseHamiltonian} and the longitudinal Hamiltonian of Eq.~\eqref{eq:longHamiltonian}:
\begin{multline}
\mathcal{H} = \left( -\frac{\hbar^2}{2m^*}\nabla^2 -e \phi(x,y)\right) \sigma_0  - i \alpha \frac{\partial}{\partial z} \sigma_y
+ E_{\textrm{Z}} \sigma_z.
\label{eq:1DHam}
\end{multline}
Assuming that the wave function has the form of a plane wave $\propto e^{ikz}$ in the longitudinal direction, and quantized transverse modes $\psi_i$ with corresponding energies $E_i$ in the transverse direction (where $i$ denotes the transverse mode number), the energies of the Hamiltonian are
\begin{equation}
E(k) = E_i + \frac{\hbar^2 k^2}{2m^*} \pm \sqrt{E_{\textrm{Z}}^2 + \alpha^2 k^2},
\label{eq:dispersion_E}
\end{equation}
yielding the dispersions of the upper and the lower spin band. Converting Eq.~\eqref{eq:dispersion_E} to momentum as a function of energy yields
\begin{multline}
k_{\pm}(E, E_i, E_{\textrm{Z}}, \alpha) = \\ \frac{1}{\sqrt{2}}\sqrt{\alpha^2 + 2(E - E_i) \pm \sqrt{\alpha^4 + 4\alpha^2(E - E_i) + 4E_{\textrm{Z}}^2}},
\label{eq:dispersion_k}
\end{multline}
where $\alpha$, $E$, $E_{\textrm{Z}}$, and $E_i$ are in units of $\hbar^2/2m^*$.
The relation between the density of states $D(E)$ and $k$ is
\begin{equation}
D(E) = \frac{1}{\pi}\frac{dk}{dE}.
\label{eq:dos}
\end{equation}
We obtain the density $n(E_i, E_Z, \alpha)$ by integrating the density of states up to the Fermi level $E_F$. The Zeeman field opens a gap of size $2E_Z$ between the upper and the lower spin band. Due to the W-shape of the lower spin band, induced by the spin-orbit interaction, we distinguish three energy regions in integrating up to $E_F$. If $E_F > E_Z$, both spin bands are occupied and the integration yields
\begin{multline}
n(E_i, E_Z, \alpha) = \frac{1}{\pi}(k_+(E_F, E_i, E_Z, \alpha) + \\
k_-(E_F, E_i, E_Z, \alpha)).
\label{eq:updensity}
\end{multline}
If $-E_Z < E_F < E_Z$, only the lower band is occupied, and the dispersion has two crossings with the Fermi level, yielding a density
\begin{equation}
n(E_i, E_Z, \alpha) = \frac{1}{\pi} k_+(E_F, E_i, E_Z, \alpha).
\label{eq:middledensity}
\end{equation}
For a nonzero spin-orbit strength, we have four crossings of the lower spin band with $E_F$ if $E_F < -E_Z$ (see Fig.~\ref{fig:mu_variation}, bottom panel). Since only the interval $k_{-} \leq k \leq k_{+}$ contributes to the density, integration of the density of states yields
\begin{multline}
n(E_i, E_Z, \alpha) = \frac{1}{\pi}(k_{+}(E_F, E_i, E_Z, \alpha) - \\
k_{-}(E_F, E_i, E_Z, \alpha)).
\label{eq:downdensity}
\end{multline}

Eqs.~\eqref{eq:updensity}, \eqref{eq:middledensity}, and \eqref{eq:downdensity} provide analytic expressions for the electron density. We use these equations to calculate the charge density of Eq.~\eqref{eq:density}.

\section{Response to the Zeeman field in the constant density limit
and for small spin-orbit}
\label{sec:constantdens_analytics}

The limit of small spin-orbit interaction and constant electron
density in the nanowire independent of Zeeman field allows for an
analytic solution the magnetic field dependence of the chemical
potential, $\mu = \mu(E_\text{Z})$. In particular, we have from
Eqs.~\eqref{eq:updensity} and \eqref{eq:downdensity} for
$\mu(E_\text{Z}=0) = \mu_0 > 0$:
\begin{multline}
\frac{2 \sqrt{2m^*}}{\pi \hbar} \sqrt{\mu_0} =
\frac{\sqrt{2 m^*}}{\pi \hbar} 
\Bigl( \sqrt{\mu+E_\text{Z}} \\+ \theta(\mu- E_\text{Z}) \sqrt{\mu - E_\text{Z}}\Bigr)\,,
\end{multline}
where $\theta$ is the Heaviside step function. This is readily solved as
\begin{equation}
\mu = \begin{cases}
\mu_0 + E_\text{Z}^2/(4 \mu_0)&\text{for $E_\text{Z} < 2 \mu_0$,}\\
4 \mu_0-E_\text{Z} &\text{for $E_\text{Z} > 2 \mu_0$.}
\end{cases}
\label{eq:constdens_mu}
\end{equation}
Hence, the chemical potential first increases with increasing $E_\text{Z}$
until the upper spin-band is completely depopulated. Then the chemical
potential decreases linearly with $E_\text{Z}$. At the cross-over point
the dependence of the chemical potential is not smooth but exhibits a kink,
also seen for example in the numerical results of Fig.~\ref{fig:mu_multiple}.

In the constant density limit we can also compute the asymptotes of
the topological phase in $\mu$-$E_\text{Z}$-space. For $E_\text{Z} \gg
\Delta$, the topological phase coincides with the chemical potential
range where only one spin subband is occupied. From Eq.~\eqref{eq:constdens_mu}
we find the two asymptotes thus as $\mu=0$ and $\mu=E_\text{Z}/2$. Hence,
in the constant density limit, the phase boundary that corresponds to
depleting the wire becomes magnetic field independent.

\section{Benchmark of nonlinear optimization methods}
\label{app:benchmarks}

\co{We use Anderson to solve Schrodinger-Poisson}

We apply the Anderson mixing scheme to solve the coupled nonlinear Schr\"odinger-Poisson equation:
\begin{equation}
\begin{cases}
\nabla^2 \phi(x,y) = -\rho(\psi_i(x,y), E_i)/\epsilon \\
\mathcal{H}[\phi(x,y)]\psi_i(x,y) = E_i\psi_i(x,y)
\end{cases}.
\label{eq:fullSP}
\end{equation}
Optimization methods find the root of the functional form of Eq.~\eqref{eq:fullSP}, as given in Eq.~\eqref{eq:minimize_func}.
As opposed to other methods, the Anderson method uses the output of last $M$ rounds as an input to the next iteration step instead of only the output of the last round. \cite{Eyert1996}
The memory of the Anderson method prevents the iteration scheme from oscillations and causes a significant speedup in computation times in comparison to other methods, and in particular the simple under-relaxation method often used in nanowire simulations.\cite{Wang2006, Chin2009}

\co{A simple test system to benchmark Anderson with other nonlinear optimization methods}

As a test system, we take a global back gate device, consisting of a hexagonal InSb nanowire on an  $\epsilon_\text{r} = 4$ dielectric layer (SiO$_2$) of thickness \SI{285}{\nm}, without a superconducting lead.
Due to the thick dielectric layer in comparison to the Majorana device, this device is more sensitive for charge oscillations (a different number of electron modes in the system between two adjacent iteration steps).
This makes the device well-suited for a performance benchmark.
We compare the Anderson method to three other nonlinear optimization methods: Broyden's First and Second method \cite{Broyden1965} and a method implementing a Newton-Krylov solver (BiCG-stab) \cite{Knoll2004}.
\begin{figure}[bt]
\centerline{\includegraphics[width=1.0\linewidth]{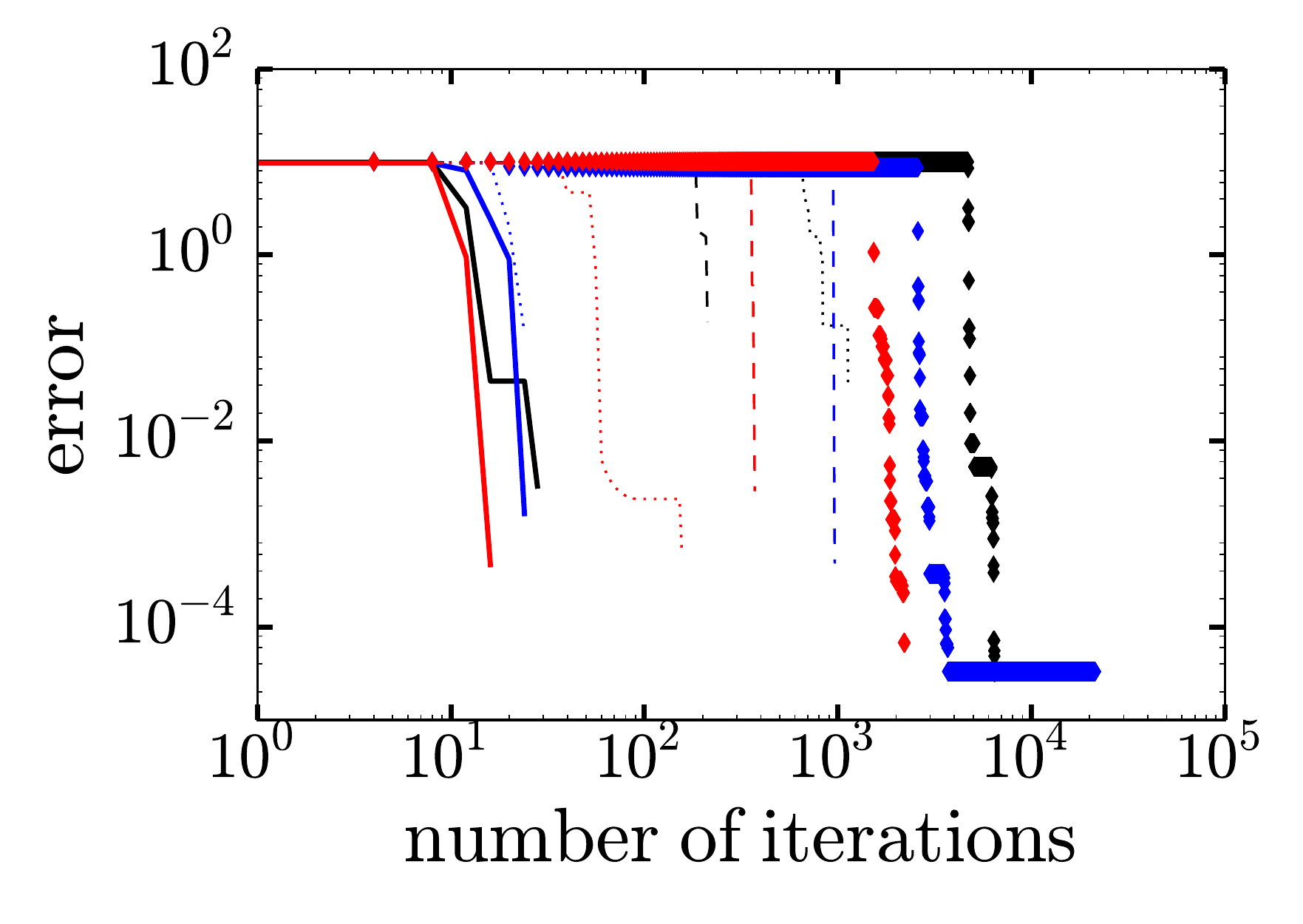}}
\caption{Benchmark of the Anderson solver (solid lines) with the First Broyden's method (dashed lines), the Second Broyden's method (dotted lines) and the BiCG-stab Newton-Krylov method (diamond markers). Black, blue and red colors correspond to a gate voltage $V_G=0.3$, 0.4, and \SI{0.5}{\volt} respectively. We show the cumulative minimum of the error.}
\label{fig:solver_benchmark}
\end{figure}

\co{Results of the performance benchmark}

Fig.~\ref{fig:solver_benchmark} shows the results. In this plot, we show the cumulative minimum of the error. Plateaus in the plot correspond to regions of error oscillations. The figure shows that the Anderson method generally converges quickly and is not affected by error oscillations.
However, the three other methods show oscillatory behavior of the error over a large range of iterations.
Both Broyden's methods perform worse than the Anderson method, but generally converge within $\sim 10^3$ iterations.
The Newton-Krylov method performs the worst, having a large region of oscillations up to $\sim 10^3 - 10^4$ iterations.
Due to its robustness against error oscillations, the Anderson method is the most suited optimization method for the Schr\"odinger-Poisson problem.
For a much thinner dielectric layer, such as the 30 nm layer in the Majorana device, the iteration number is typically $\sim 10^1$ for all four tested optimization methods.

\co{Motivation to choose a direct approach}

In our approach, we choose not to use a predictor-corrector approach\cite{Trellakis1997, Curatola2003} that can also be used together with a more advanced nonlinear solver such as the Anderson method.\cite{Wang2009} The advantage of the direct approach used here is its simplicity, without a significant compromise in stability and efficiency.

\bibliographystyle{apsrev4-1}
\bibliography{estat}

\end{document}